\documentclass[english,aps]{revtex4}
\usepackage[T1]{fontenc}
\usepackage[latin9]{inputenc}
\usepackage{refstyle}
\usepackage{amsmath}
\usepackage{amssymb}
\usepackage{graphicx}
\usepackage{esint}

\makeatletter

\RS@ifundefined{subref}
  {\def\RSsubtxt{section~}\newref{sub}{name = \RSsubtxt}}
  {}
\RS@ifundefined{thmref}
  {\def\RSthmtxt{theorem~}\newref{thm}{name = \RSthmtxt}}
  {}
\RS@ifundefined{lemref}
  {\def\RSlemtxt{lemma~}\newref{lem}{name = \RSlemtxt}}
  {}

\@ifundefined{textcolor}{}
{%
 \definecolor{BLACK}{gray}{0}
 \definecolor{WHITE}{gray}{1}
 \definecolor{RED}{rgb}{1,0,0}
 \definecolor{GREEN}{rgb}{0,1,0}
 \definecolor{BLUE}{rgb}{0,0,1}
 \definecolor{CYAN}{cmyk}{1,0,0,0}
 \definecolor{MAGENTA}{cmyk}{0,1,0,0}
 \definecolor{YELLOW}{cmyk}{0,0,1,0}
}

\makeatother

\usepackage{babel}
\begin{document}


\title{Variational theory of soliplasmon resonances}

\author{A. Ferrando$^{1}$ }
\email{albert.ferrando@uv.es}
\homepage{http://www.intertech.upv.es/members_photonics.htm}
\affiliation{
$^{1}$Departament d'\`{O}ptica, Interdisciplinary Modeling Group InterTech,
Universitat de Val\`{e}ncia, Dr. Moliner 50, 46100
Burjassot (Val\`{e}ncia), Spain}

\author{C. Mili\'{a}n$^{2,3}$}
\affiliation{
$^{2}$Instituto de Instrumentaci\'on para Imagen Molecular (I3M), InterTech,
Universitat Polit\`ecnica de Val\`encia, Camino de Vera
S/N 46022 Valencia, Spain}

\author{D. V. Skryabin$^{3}$}
\affiliation{
$^{3}$Centre for Photonics and Photonic Materials, Department of Physics,
University of Bath, Bath BA2 7AY, United Kingdom\\
}



\begin{abstract}
We present a first-principles derivation of the variational equations describing the dynamics of the interaction of a spatial soliton and a surface plasmon polariton (SPP) propagating along a metal/dielectric interface. The variational {\em ansatz} is based on the existence of solutions exhibiting differentiated and spatially resolvable localized soliton and SPP components. These states, referred to as {\em soliplasmons}, can be physically understood as bound states of a soliton and a SPP. Their respective dispersion relations permit the existence of a resonant interaction between them, as pointed out in \cite{Bliokh_PRA_2009}. The existence of soliplasmon states and their interesting nonlinear resonant behavior has been validated already by full-vector simulations of the nonlinear Maxwell's equations, as reported in \cite{Milian_OL_2012}. Here, we provide the theoretical demonstration of the  nonlinear resonator model previously introduced in our previous work  and analyze all the approximations needed to obtain it. We also provide some extensions of the model to improve its applicability.
\end{abstract}

\maketitle

\section{Introduction}

The idea that a spatial soliton and a surface plasmon polariton (SPP) can couple to form the soliton-plasmon, or \textit{soliplasmon}, bound state was originally proposed in Ref. \cite{Bliokh_PRA_2009}. These authors proposed by means of a sharp physical intuition that the bound system should obey in certain limit a nonlinearly coupled oscillator model with a peculiar nonlinearity, originated by the soliton tail at the metal interface, which was considered as the driving mechanism for the coupling and the dynamical features of the soliplasmon system. Despite the amount of undetermined coefficients in the model, relevant qualitative predictions of the soliplasmon properties were made. These nonlinear modes where later numerically demonstrated to exist in the context of the general vectorial nonlinear Maxwell equations and the nonlinear oscillator model (found with asymmetric coupling) was presented with fully determined coefficients \cite{Milian_OL_2012}. However, that oscillator model was never proved.

Although the soliplasmon proposal is relatively recent, monochromatic surface waves in the vicinity of the interface between a nonlinear dielectric and a metal were extensively studied during the decade of the 80's by many authors (see, to cite only a few, \cite{Agranovich_JETP_1980,Tomlinson_OL_1980,Akhmediev_JETP_1982,Yu1983a,Leung_PRB_1985,Mihalache_PS_1984,Stegeman1985a,Ariyasu_JAP_1985}). Amongst those theoretical results, corrections to the profile and wavenumber of i) SPP's due to nonlinearity and ii) of the spatial solitons due to the presence of a metallic interface far from its core, were presented.  At that time, none of the studies, to the best of our knowledge, considered simultaneously a spatial soliton and a SPP; the two component wave referred to as soliplasmon. Recently, the model presented in \cite{Ariyasu_JAP_1985} has been extended in a first attempt to describe soliplasmons in the challenging and more realistic 2D geometry, consisting of several interfaces \cite{Walasik_OL_2012}. So far the 2D symmetric like modes are reported, but the antisymmetric ones, which in principle require less power in the soliton component \cite{Milian_OL_2012}, remain unfound. On the other hand, the role of nonlinear vector effects, generated by strong gradients of the effective nonlinearly induced  refractive index, can be of relevance for understanding soliplasmon modes near resonance or with high plasmonic component, such as in metals directly attached to the nonlinear medium, so standard scalar soliton solutions cannot be appropriate when they are too close to the interface and a full-vector soliton solution is required \cite{Ciattoni2005a}.

In this paper, we present the detailed derivation of the nonlinear oscillator model from first principles. We focus on 1D soliplasmons in two geometries of interest, namely the metal/Kerr (MK) \cite{Milian_OL_2012} and the metal/dielectric/Kerr (MDK) \cite{Bliokh_PRA_2009} interfaces (by \textit{dielectric} we implicitly mean a linear dielectric and by a Kerr  dielectric with cubic nonlinearity).  The paper is organized as follows: in  Section II, we motivate and present our variational {\em ansatz}  and the main features of soliton-plasmon coupling in the general context of nonlinear Maxwell's equations; in Section III, we focus on the soliton component of the soliplasmon {\em ansatz} in the case of total decoupling and we obtain the dynamical equation for the soliton variational parameter; in Section IV, we obtain the equation for a general nonlinear plasmon stationary mode (decoupled from soliton) and construct the dynamical equation for the plasmon variational parameter. Finally, in Section V, we obtain the nonlinear resonator model for the coupled system formed by a nonlinear plasmon and a soliton in the weak coupling approximation. In Section VI, we pay attention to the cases of a MK and MDK geometries previous mentioned recovering and demonstrating the model presented in Refs.\cite{Bliokh_PRA_2009,Milian_OL_2012}. In all sections, we stress the role played  by the different  approximations used in our derivation.  The relevance and differences of this model with respect other approaches are especially emphasized in the conclusions.

\section{Variational \emph{ansatz} for nonlinear Maxwell's equations\label{sec:variational_ansatz_NL_maxwell_eqs}}

We consider the wave equation for the electric component of a monochromatic
EM field describing its propagation in an optical media characterized
by inhomogeneous linear and nonlinear (Kerr) susceptibilities. We
take the most general form of these equations derived from Maxwell's
equations without assuming neither the scalar nor the paraxial approximations.
Thus, our starting point is:

\begin{equation}
\nabla^{2}\mathbf{E}-\nabla\left(\nabla\cdot\mathbf{E}\right)=-k_{0}^{2}\mathbf{D}\left(\mathbf{E}\right)=-k_{0}^{2}\varepsilon_{L}\mathbf{E}-k_{0}^{2}\mathbf{P}_{\mathrm{NL}}\left(\mathbf{E}\right)\label{eq:NL_vector_wave_eq}
\end{equation}
where $\varepsilon_{L}(\mathbf{x})\equiv1+\chi^{(1)}(\mathbf{x)}$
is the in-homogenous linear dielectric function and 
\[
\mathbf{P}_{\mathrm{NL}}\left(\mathbf{E}\right)=\chi^{(3)}\left(\mathbf{E}\cdot\mathbf{E}^{*}\right)\mathbf{E}+\bar{\chi}^{(3)}\left(\mathbf{E}\cdot\mathbf{E}\right)\mathbf{E}^{*},
\]

 is the Kerr nonlinear polarization, where $\chi^{(3)}(\mathbf{x})$
and $\bar{\chi}^{(3)}(\mathbf{x})$ are the in-homogeneous third order
susceptibilities associated to the Kerr effect. All functions have
an implicit, but undisplayed, dependence on the frequency of the nonlinear
wave $\omega=ck_{0}$.

\begin{figure}
\hfill{}\includegraphics[scale=0.4]{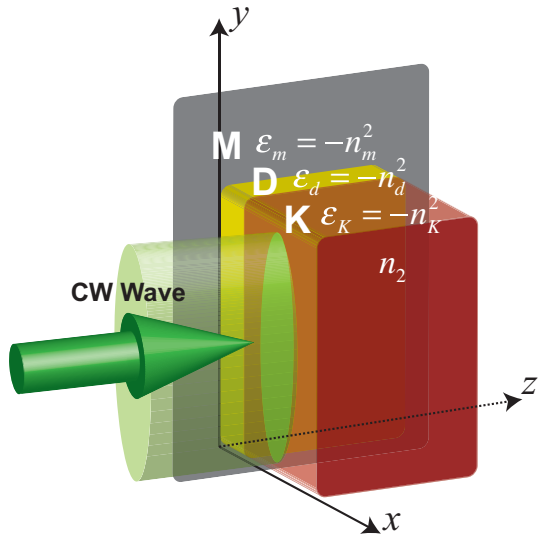}\hfill{}

\caption{Parallel illumination of a metal/dielectric interface from a Kerr
medium.\label{fig:MDK}}
\end{figure}

Our goal is to find a simplified, although physically meaningful,
model describing the existence of soliton-plasmon resonances, or \emph{soliplasmons},
obtained by the interaction of a spatial soliton of the Kerr medium
with a Surface Plasmon Polariton (SPP) on a metal/dielectric interface.
We consider a simple configuration consisting in a spatial soliton
moving (within a Kerr medium) in parallel to a metal/dielectric interface
interacting with a SPP propagating on it (see Fig.\ref{fig:MDK}).
It has been proven that stationary nonlinear states of this system
in the form of soliton-plasmon resonances indeed exists as a solution
of Eq.(\ref{eq:NL_vector_wave_eq}) for this type of configuration
\cite{Milian_OL_2012}. They show a structure in which the soliton and
plasmon components are clearly distinguishable, which supports to
adopt the following variational \emph{ansatz:
\begin{equation}
\mathbf{E}=\mathbf{E}_{\mathrm{np}}\left[\left\{ A_{i}(z)\right\} _{i=1}^{N}\right]+\mathbf{E}_{s}\left[\left\{ C_{i}(z)\right\} _{i=1}^{N}\right],\label{eq:general_ansatz}
\end{equation}
}in which, in principle, we do not explicitly display the dependence
on the variational parameters so that a multi-parametric dependence
can be assumed. We will particularize these parameters later. We also
allow the SPP to behave nonlinearly by the effect of the Kerr nonlinearity
on its own propagation \cite{Agranovich_JETP_1980,Davoyan_OE_2009}
thus giving rise to what we call a nonlinear plasmon. We introduce
the general \emph{ansatz} (\ref{eq:general_ansatz}) in Eq. (\ref{eq:NL_vector_wave_eq})
to get:
\begin{eqnarray}
\frac{\partial^{2}\mathbf{E}_{\mathrm{np}}}{\partial z^{2}}+L_{0}\mathbf{E}_{\mathrm{np}}-\nabla\left(\nabla\cdot\mathbf{E}_{\mathrm{np}}\right)+\frac{\partial^{2}\mathbf{E}_{s}}{\partial z^{2}}+L_{0}\mathbf{E}_{s} & =\nonumber \\
\nonumber \\
=-k_{0}^{2}\mathbf{P}_{\mathrm{NL}}\left(\mathbf{E}_{\mathrm{np}}\right)-k_{0}^{2}\mathbf{P}_{\mathrm{NL}}\left(\mathbf{E}_{\mathrm{s}}\right)-k_{0}^{2}\mathbf{Q}_{\mathrm{K}}\left(\mathbf{E}_{\mathrm{np}},\mathbf{E}_{\mathrm{s}}\right),\label{eq:wave_eq_Enp+Es}
\end{eqnarray}
where we have taken into account that the soliton is essentially a
scalar solution so that $\nabla\cdot\mathbf{E}_{s}\approx0$. We have
also defined the differential operator $L_{0}$ as $L_{0}\equiv\nabla_{t}^{2}+k_{0}^{2}\varepsilon_{L}$
where $\nabla_{t}=\left(\partial_{x},\partial_{y}\right)$ is the
transverse gradient operator. The term $\mathbf{Q}_{\mathrm{K}}\left(\mathbf{E}_{\mathrm{np}},\mathbf{E}_{\mathrm{s}}\right)$
includes all nonlinear Kerr terms coupling the SPP and soliton components:
\begin{equation}
\mathbf{Q}_{\mathrm{K}}\left(\mathbf{E}_{\mathrm{np}},\mathbf{E}_{\mathrm{s}}\right)\equiv\mathbf{P}_{\mathrm{NL}}\left(\mathbf{E}_{\mathrm{np}}+\mathbf{E}_{s}\right)-\mathbf{P}_{\mathrm{NL}}\left(\mathbf{E}_{\mathrm{np}}\right)-\mathbf{P}_{\mathrm{NL}}\left(\mathbf{E}_{\mathrm{s}}\right).\label{eq:NL_Kerr_coupling}
\end{equation}
The SPP and soliton fields have different features. The SPP, as a
surface wave and even when it behaves nonlinearly, can only exist
bounded to the metal/dielectric interface, its nature being purely
vectorial. On the other hand, the soliton can move freely in the Kerr
medium with no restriction and its existence is due to the Kerr nonlinearity
and it is essentially a scalar wave. It is clear from physical arguments
and from results in Ref.\cite{Milian_OL_2012} that the coupling between
these two entities vanishes as we move the soliton far way from the
interface. In such a case, the overlapping of $\mathbf{E}_{\mathrm{np}}$
and $\mathbf{E}_{s}$ tends to zero and, thus, $\mathbf{Q}_{\mathrm{K}}\rightarrow0$.
In the limiting case in which they are infinitely far away, Eq.(\ref{eq:NL_vector_wave_eq})
gives rise to two independent equations for $\mathbf{E}_{\mathrm{np}}$
and $\mathbf{E}_{s}$:
\begin{equation}
\frac{\partial^{2}\mathbf{E}_{\mathrm{np}}}{\partial z^{2}}+L_{0}\mathbf{E}_{\mathrm{np}}-\nabla\left(\nabla\cdot\mathbf{E}_{\mathrm{np}}\right)=-k_{0}^{2}\mathbf{P}_{\mathrm{NL}}\left(\mathbf{E}_{\mathrm{np}}\right)\label{eq:NL_plasmon_wave_eq}
\end{equation}
 and
\begin{equation}
\frac{\partial^{2}\mathbf{E}_{s}}{\partial z^{2}}+L_{0}\mathbf{E}_{s}=-k_{0}^{2}\mathbf{P}_{\mathrm{NL}}\left(\mathbf{E}_{\mathrm{s}}\right).\label{eq:soliton_wave_eq}
\end{equation}
\[
\]
Thus we can consider the action of $\mathbf{Q}_{\mathrm{K}}$ as a
perturbation that couples the solutions of these two independent equations.
For this reason, in our variational approach we will first consider
the situation in which the two previous equations are satisfied separately.
We will write the variational equations for both components separately
and then we introduce the first order correction originated by the
coupling term $\mathbf{Q}_{\mathrm{K}}$ .

\section{Variational equation for the soliton\label{sec:variational_eq_soliton}}

We start by considering the equation of an uncoupled soliton field
satisfying Eq.(\ref{eq:soliton_wave_eq}). We consider the simplest
case by choosing the fundamental soliton solution as a variational
\emph{ansatz} and by taking a single variational parameter, namely,
the soliton amplitude. In the planar geometry under consideration,
as given by Fig.\ref{fig:MDK}, we consider a $x$ and $z$ dependence
only, so that the fundamental soliton corresponds to that of the 1D
Helmholtz equation. This \emph{stationary }solution has the \emph{sech
}form:
\begin{equation}
\mathbf{\bar{E}}_{\mathrm{s}}(x,z)=\mathbf{u}C\mathrm{sech}\left[\sqrt{\frac{\gamma}{2}}k_{0}C\left(x-a\right)\right]e^{i\beta_{s}z}\equiv\mathbf{u}\bar{E}_{s}(x;C)e^{i\beta_{s}z},\,\,\,\,\,\, C\in\mathbb{R}^{+},\,\,\, a\text{\ensuremath{\in}}\mathbb{R}^{+},\label{eq:helmholtz_soliton_solution}
\end{equation}
where $\gamma=\chi^{(3)}+\bar{\chi}^{(3)}$, $\mathbf{u}$ is a real
unitary vector and the propagation constant of the soliton is given
by 
\[
\beta_{s}^{2}=k_{0}^{2}\left(\varepsilon_{\mathrm{K}}+\frac{\gamma}{2}C^{2}\right),
\]
$\varepsilon_{\mathrm{K}}$ being the linear dielectric constant of
the Kerr medium. It is easy to check that the expression (\ref{eq:helmholtz_soliton_solution})
verifies:
\begin{equation}
\left[\nabla_{t}^{2}+k_{0}^{2}\left(\varepsilon_{\mathrm{K}}+\gamma\left|\mathbf{\bar{E}}_{\mathrm{s}}\right|^{2}\right)\right]\bar{\mathbf{E}}_{\mathrm{s}}=\beta_{s}^{2}\mathbf{\bar{E}}_{\mathrm{s}},\label{eq:stationary_soliton_eq}
\end{equation}
or, equivalently, Eq.(\ref{eq:soliton_wave_eq}) for a stationary
solution in which $\mathbf{P}_{\mathrm{NL}}=\gamma\left|\mathbf{\bar{E}}_{\mathrm{s}}\right|^{2}\mathbf{\bar{E}}_{\mathrm{s}}$. 

Now, we make the \emph{ansatz} for the soliton component. As shown
in Ref.\cite{Milian_OL_2012}, for quasi-stationary evolution the dynamics
of the soliton position $a$ is much slower than that of the amplitude.
For this reason, we promote the soliton amplitude $C$ to the category
of the only variational parameter for the case under consideration.
Therefore, we establish the \emph{ansatz} as follows:
\begin{equation}
\mathbf{\bar{E}}_{s}(x,z)\rightarrow\mathbf{E}_{s}(x,z)=\mathbf{u}C(z)\mathrm{sech}\left[\sqrt{\frac{\gamma}{2}}k_{0}\left|C(z)\right|\left(x-a\right)\right]\,\,\,\,\,\, C(z)\in\mathbb{C}.\label{eq:soliton_ansatz}
\end{equation}
Our aim now is to find the dynamical equation for the variational
parameter $C(z)$. We immediately recognize that the previous ansatz
admits the following useful decomposition (taking into account that
we can write $C(z)=\left|C(z)\right|\exp\left[i\varphi_{s}(z)\right]$):
\[
\mathbf{E}_{s}(x,z)=\mathbf{u}e^{i\varphi_{s}(z)}\bar{E}_{s}(x;|C(z)|),
\]
where $\bar{E}_{s}(x;|C(z)|)$ is the stationary soliton solution
characterized by the real and positive amplitude $ $$|C(z)|$ and,
consequently, by a propagation constant $\beta_{s}=k_{0}\left(\varepsilon_{\mathrm{K}}+\frac{\gamma}{2}|C\left(z\right)|^{2}\right)^{1/2}$.
Since we are interested in quasi-stationary evolution, we will assume
that the dynamics of $ $$|C(z)|$ is much slower than that of the
corresponding phase $\varphi_{s}(z)$, so that, $d|C|/dz\ll d\varphi_{s}/dz$.
This assumption is supported by numerical simulations \cite{Milian_OL_2012}.
We substitute now the \emph{ansatz} Eq.(\ref{eq:soliton_ansatz})
into the soliton wave equation (\ref{eq:soliton_wave_eq}):
\[
\frac{\partial^{2}\mathbf{E}_{s}}{\partial z^{2}}+L_{0}\mathbf{E}_{s}=-k_{0}^{2}\left[\left(\chi^{(3)}+\bar{\chi}^{(3)}\right)\left|\mathbf{E}_{s}\right|^{2}\right]\mathbf{E}_{s},
\]
which can be written as
\begin{equation}
\frac{\partial^{2}\mathbf{E}_{s}}{\partial z^{2}}+\left[\nabla_{t}^{2}+k_{0}^{2}\left(\varepsilon_{\mathrm{K}}+\gamma\left|\mathbf{E}_{\mathrm{s}}\right|^{2}\right)\right]\mathbf{E}_{\mathrm{s}}=0.\label{eq:variational_eq_soliton}
\end{equation}
But, since according to our \emph{ansatz,} $\mathbf{E}_{s}(x,z)=\mathbf{u}\exp\left[i\varphi_{s}(z)\right]\bar{E}_{s}(x,z)\equiv\mathbf{u}E_{s}(x,z)$,
we can see that the differential operator in brackets only acts on
the stationary solution $\bar{E}_{s}$. Since the latter field, on
the other hand, satisfies the stationary equation (\ref{eq:stationary_soliton_eq}),
this means that the variational field $\mathbf{E}_{s}$ also fulfills
\begin{equation}
\left[\nabla_{t}^{2}+k_{0}^{2}\left(\varepsilon_{\mathrm{K}}+\gamma\left|\mathbf{E}_{\mathrm{s}}\right|^{2}\right)\right]\mathbf{E}_{\mathrm{s}}=\beta_{s}^{2}\mathbf{E}_{\mathrm{s}},\label{eq:NL_eigenvalue_eq_varitional_soliton}
\end{equation}
and, therefore, the previous equation becomes:
\[
\frac{\partial^{2}E_{s}\left(x;C(z)\right)}{\partial z^{2}}+\beta_{s}^{2}E_{\mathrm{s}}\left(x;C(z)\right)=0.
\]
At this point, we introduce another useful decomposition for the variational
field (\ref{eq:soliton_ansatz}): $ $$E_{s}(x,z)=C(z)f_{s}(x;|C(z)|)$,
where $f_{s}=\mathrm{sech}\left[\sqrt{\frac{\gamma}{2}}k_{0}\left|C(z)\right|\left(x-a\right)\right].$
This decomposition permits to write the equation above as:
\[
\frac{\partial^{2}}{\partial z^{2}}\left[C(z)f_{s}(x;|C(z)|\right]+\beta_{s}^{2}\left[C(z)f_{s}(x;|C(z)|\right]=0.
\]
In order to obtain a dynamical equation for $C(z)$, we need to project
the previous equation with respect to a suitable function. A natural
choice for the projection function is the stationary soliton solution
at the initial propagation point, i.e., $\bar{E}_{s}(0)=C(0)f_{s}(x;|C(0)|)$.
Since $C(0)$ is constant, it disappears in the projection process,
so that, after projection we have:
\[
\frac{d^{2}}{dz^{2}}\left[C(z)\int_{\mathbb{R}}f_{s}\left(x,0\right)f_{s}(x;|C(z)|)\right]+\beta_{s}^{2}\left[C(z)\int_{\mathbb{R}}f_{s}\left(x,0\right)f_{s}(x;|C(z)|)\right]=0.
\]
We introduce now the notation $N_{s}\equiv\int_{\mathbb{R}}f_{s}(0)f_{s}(z)$
and we immediately recognize that $ $the dependence of $N_{s}$ on
$z$ comes exclusively from its dependence on the modulus of the soliton
amplitude $|C(z)|$. However, due to the quasi-stationary approximation
$d|C|/dz\ll d\varphi_{s}/dz$, 
%
%
\begin{equation}
\frac{d^{2}}{dz^{2}}\left[C(z)N_{s}\left(|C(z)|\right)\right]\approx\left(\frac{d^{2}C(z)}{dz^{2}}\right)N_{s}\left(|C(z)|\right).\label{eq:quasi_stationary_approx_for_C}
\end{equation}
Therefore, the variational equation for the soliton parameter takes
the simple form:
\[
N_{s}\left[\frac{d^{2}}{dz^{2}}C(z)+\beta_{s}^{2}\left(|C|\right)C(z)\right]=0.
\]

\section{Variational equation for the nonlinear plasmon\label{sec:variational_equation_for_NL_plasmon}}

In the case of the nonlinear plasmon component, we follow a similar
procedure as for the soliton case. However, equations are here more
cumbersome to analyze since we have to deal with the vectorial part
of the differential operator in the wave equation (\ref{eq:NL_plasmon_wave_eq}).
Another difference is that, since the SPP is a surface wave, the linear
dielectric function is no longer a constant, as for the soliton equation,
but rather a function defining the dielectric/metal interface:
\begin{equation}
\varepsilon_{L}(x)=\varepsilon_{p}(x)\equiv\begin{cases}
\varepsilon_{m} & \mathrm{if\,\,}x\le0\\
\varepsilon_{d} & \mathrm{if\,\,}x>0.
\end{cases}\label{eq:dielectric_function_MD_interface}
\end{equation}
A nonlinear plasmon is a stationary solution of Eq.(\ref{eq:NL_plasmon_wave_eq})
of the form:
\[
\mathbf{\bar{E}}_{\mathrm{np}}(x,z)=\mathbf{e}_{\mathrm{np}}(x)e^{i\beta_{\mathrm{np}}z}=\left(\begin{array}{c}
\mathbf{e}_{\mathrm{np}t}(x)\\
e_{\mathrm{np}z}(x)
\end{array}\right)e^{i\beta_{\mathrm{np}}z},
\]
where $\mathbf{e}_{\mathrm{np}t}=\left(e_{\mathrm{np}x},e_{\mathrm{np}y}\right)^{\top}$
stands for the transverse components of the electric field. We will
consider that the nonlinear plasmon stationary solution is a conservative
soliton. This means that for the stationary solution we will assume
that the system has no losses, neither linear nor nonlinear, so that
$\varepsilon_{L}$ will be a real function. For the same reason, we
will take real nonlinear susceptibilities ($\chi^{(3)},\bar{\chi}^{(3)}\in\mathbb{R}$
). The complex character of a realistic $ $$\varepsilon_{L}$ will
be taken into account when we set the dynamical equations for the
variational solution.

According to Eq.(\ref{eq:NL_plasmon_wave_eq}), the transverse components
of the nonlinear plasmon solution verify:
\begin{eqnarray}
-\beta_{\mathrm{np}}^{2}\mathbf{e}_{\mathrm{np}t}+\nabla^{2}\mathbf{e}_{\mathrm{np}t}-\nabla_{t}\left(i\beta_{\mathrm{np}}e_{\mathrm{np}z}+\nabla_{t}\cdot\mathbf{e}_{\mathrm{np}t}\right) & =\nonumber \\
 &  & \hspace{-4cm}-k_{0}^{2}\varepsilon_{p}\mathbf{e}_{\mathrm{np}t}-k_{0}^{2}\left[\chi^{(3)}\left(\mathbf{e}_{\mathrm{np}}\cdot\mathbf{e}_{\mathrm{np}}^{*}\right)\mathbf{e}_{\mathrm{np}t}+\bar{\chi}^{(3)}\left(\mathbf{e}_{\mathrm{np}}\cdot\mathbf{e}_{\mathrm{np}}\right)\mathbf{e}_{\mathrm{np}t}^{*}\right].\label{eq:NL_plasmon_eq_transverse}
\end{eqnarray}
In the linear case, the transverse components of the electric field
corresponding to the eigenmodes of an axially-invariant system can
be chosen to be real functions ($ $$\mathbf{e}_{\mathrm{np}t}=\mathbf{e}_{\mathrm{np}t}^{*}$)
whereas the axial ones are pure imaginary ($e_{\mathrm{np}z}=-e_{\mathrm{np}z}^{*}$)
\cite{Snyder_book_1983}. As we will see next, this choice is also consistent
in the nonlinear vector case. Assuming these properties for the electric
field, the transverse component of the nonlinear polarization term
associated to the previous equation can be written as 
\begin{eqnarray*}
\mathbf{P}_{\mathrm{NL}t} & = & \left[\chi^{(3)}\left(\mathbf{e}_{\mathrm{np}}\cdot\mathbf{e}_{\mathrm{np}}^{*}\right)\mathbf{e}_{\mathrm{np}t}+\bar{\chi}^{(3)}\left(\mathbf{e}_{\mathrm{np}}\cdot\mathbf{e}_{\mathrm{np}}\right)\mathbf{e}_{\mathrm{np}t}^{*}\right]e^{i\beta_{\mathrm{np}}z}\\
 & = & \left[\chi^{(3)}\left(\mathbf{e}_{\mathrm{np}t}\cdot\mathbf{e}_{\mathrm{np}t}-e_{\mathrm{np}z}^{2}\right)\mathbf{e}_{\mathrm{np}t}+\bar{\chi}^{(3)}\left(\mathbf{e}_{\mathrm{np}t}\cdot\mathbf{e}_{\mathrm{np}t}+e_{\mathrm{np}z}^{2}\right)\mathbf{e}_{\mathrm{np}t}\right]e^{i\beta_{\mathrm{np}}z}\\
 & = & \left[\gamma\left|\mathbf{e}_{\mathrm{np}t}\right|^{2}-\bar{\gamma}e_{\mathrm{np}z}^{2}\right]\mathbf{e}_{\mathrm{np}t}e^{i\beta_{\mathrm{np}}z},
\end{eqnarray*}

where $\gamma\equiv\chi^{(3)}+\bar{\chi}^{(3)}$ and $\bar{\gamma}\equiv\chi^{(3)}-\bar{\chi}^{(3)}$.
An analogous calculation leads to the following relation for the axial
component
\[
P_{\mathrm{NL}z}=\left[\bar{\gamma}\left|\mathbf{e}_{\mathrm{np}t}\right|^{2}-\gamma e_{\mathrm{np}z}^{2}\right]e_{\mathrm{np}z}e^{i\beta_{\mathrm{np}}z}.
\]

The total displacement vector $\mathbf{D}=\varepsilon \mathbf{E} +\mathbf{P}_{\mathrm{NL}}$
takes then the form
\begin{eqnarray}
\mathbf{D}_{t} & = & \left[\varepsilon_{\mathrm{L}}+\left(\gamma\left|\mathbf{e}_{\mathrm{np}t}\right|^{2}-\bar{\gamma}e_{\mathrm{np}z}^{2}\right)\right]\mathbf{e}_{\mathrm{np}t}e^{i\beta_{\mathrm{np}}z}\equiv\varepsilon_{\mathrm{np}}\mathbf{e}_{\mathrm{np}t}e^{i\beta_{\mathrm{np}}z}\nonumber \\
D_{z} & = & \left[\varepsilon_{\mathrm{L}}+\left(\bar{\gamma}\left|\mathbf{e}_{\mathrm{np}t}\right|^{2}-\gamma e_{\mathrm{np}z}^{2}\right)\right]e_{\mathrm{np}z}e^{i\beta_{\mathrm{np}}z}\equiv\bar{\varepsilon}_{\mathrm{np}}e_{\mathrm{np}z}e^{i\beta_{\mathrm{np}}z}.\label{eq:total_polarization}
\end{eqnarray}

On the other hand, due to the mathematical identity:
\[
\nabla\left[\nabla^{2}\mathbf{E}-\nabla\left(\nabla\cdot\mathbf{E}\right)\right]\equiv0,
\]

it is identically verified from the nonlinear vector wave equation
(\ref{eq:NL_vector_wave_eq}) that
\[
\nabla\cdot\mathbf{D}=0,
\]
which imposes a constraint between the axial and transverse components
of the stationary solutions:
\begin{equation}
\nabla_{t}\cdot\left(\varepsilon_{\mathrm{np}}\mathbf{e}_{\mathrm{np}t}\right)+i\beta_{\mathrm{np}}\bar{\varepsilon}_{\mathrm{np}}e_{\mathrm{np}z}=0\Rightarrow e_{\mathrm{np}z}=\frac{i}{\beta_{\mathrm{np}}}\frac{1}{\bar{\varepsilon}_{\mathrm{np}}}\nabla_{t}\cdot\left(\varepsilon_{\mathrm{np}}\mathbf{e}_{\mathrm{np}t}\right).\label{eq:constraint}
\end{equation}
Despite its form, the previous equation does not provide an explicit
expression of the axial component in terms of the transverse ones.
The reason is the dependence of both $\varepsilon_{\mathrm{np}}$
and $\bar{\varepsilon}_{\mathrm{np}}$ on $e_{\mathrm{np}z}$ as well.
In the most general case, the solution of the nonlinear vector problem
requires to solve the transverse equation (\ref{eq:NL_plasmon_eq_transverse})
$ $ along with the constraint (\ref{eq:constraint}) in a self-consistent
manner. The form of the constraint (\ref{eq:constraint}) also demonstrates
the consistency of the assumptions with respect the real character
of transverse components and the pure imaginary condition for the
axial one. Indeed, this constraint shows that if $\mathbf{e}_{\mathrm{np}t}\in\mathbb{R}$
automatically $e_{\mathrm{np}z}$ becomes a pure imaginary function.
This is so since we are considering $\varepsilon_{L}$, $\chi^{(3)}$
and $\bar{\chi}^{(3)}$ to be real, so that $ $$\varepsilon_{\mathrm{np}}$
and $\bar{\varepsilon}_{\mathrm{np}}$ also are.


In many situations, despite the stationary eigenmodes
have an hybrid nature, the axial component is commonly remarkably
smaller than the transverse one. So that, we can reasonably consider
in many circumstances that $|e_{\mathrm{np}z}|\ll|\mathbf{e}_{\mathrm{np}t}|$.
We refer to this condition as the \emph{quasi-transverse approximation}
and, in practice, it will implemented by neglecting terms which are
second order in the axial component, i.e., $O(e_{\mathrm{np}z}^{2})\rightarrow0$.

When the quasi-transverse approximation
is considered, the nonlinear vector eigenmode can be described by
two nonlinear effective functions depending only on transverse components:
\[
\varepsilon_{\mathrm{np}}\approx\varepsilon_{L}+\gamma|\mathbf{e}_{\mathrm{np}t}|^{2}, \hspace{2cm}
\bar{\varepsilon}_{\mathrm{np}}\approx\varepsilon_{L}+\bar{\gamma}|\mathbf{e}_{\mathrm{np}t}|^{2} .
\]
In this approximation, transverse and axial components decouple in
the equation for $\mathbf{e}_{\mathrm{np}t}$ Eq.(\ref{eq:NL_plasmon_eq_transverse}) 
since, according to the constraint (\ref{eq:constraint}), 
it is verified that 
\begin{equation}
i\beta_{\mathrm{np}}e_{\mathrm{np}z}\approx -\frac{\varepsilon_{\mathrm{np}}}{\bar{\varepsilon}_{\mathrm{np}}}\nabla_{t}\cdot\mathbf{e}_{\mathrm{np}t}-\frac{\nabla_{t}\varepsilon_{\mathrm{np}}}{\bar{\varepsilon}_{\mathrm{np}}}\cdot\mathbf{e}_{\mathrm{np}t},\label{eq:contraint_transverse_approx}
\end{equation}

The previous equation permits, after substitution into Eq.(\ref{eq:NL_plasmon_eq_transverse}),
to eliminate the axial component completely from the transverse equation.
\begin{equation}
\left(\nabla_{t}^{2}+k_{0}^{2}\varepsilon_{\mathrm{np}}\right)\mathbf{e}_{\mathrm{np}t}+\nabla_{t}\left(\mathbf{F}_{\mathrm{np}t}\cdot\mathbf{e}_{\mathrm{np}t}\right)=\beta_{\mathrm{np}}^{2}\mathbf{e}_{\mathrm{np}t},\label{eq:NL_equation_only_transverse}
\end{equation}
where $\mathbf{F}_{\mathrm{np}t}\equiv\bar{\varepsilon}_{\mathrm{np}}^{-1}\nabla_{t}\varepsilon_{\mathrm{np}}+\delta \nabla_t$, $\delta=(\bar{\varepsilon}_{\mathrm{np}}-{\varepsilon}_{\mathrm{np}})/\bar{\varepsilon}_{\mathrm{np}}$ being the nonlinearly-induced anisotropy function.
Note that due to the quasi-transverse approximation, $\varepsilon_{\mathrm{np}}$
and $\mathbf{F}_{\mathrm{np}t}$ depend on transverse components exclusively.
Despite this fact, it is important to remark that in this approximation
axial components are \emph{nonzero. }They are simply decoupled from
the transverse ones. They can be obtained in a simple way from the
constraint (\ref{eq:contraint_transverse_approx}) once the problem
have been solved for the transverse components. Unlike for the general
case, the constraint becomes now an explicit expression of $e_{\mathrm{np}z}$
as a function of $\mathbf{e}_{\mathrm{np}t}$. From the computational
point of view, the decoupling of axial and transverse components considerably
simplifies the calculation of the stationary solution since, then,
the simultaneous self-consistent resolution of the transverse equation
and the constraint can be circumvented.

Up to now, all the analysis is valid for a general linear dielectric
function profile $\varepsilon_{L}$ which does not need to be necessarily
that of a SPP on a metal/dielectric interface. This means that results
can be applied to arbitrary axially-invariant structures even in 2D.
However, since we are interested in the case of a SPP on a planar
structure (Fig.\ref{fig:MDK}), we will assume that we deal with a
1D nonlinear plasmon in a TM configuration. The corresponding electric
field has then the form $\mathbf{e}_{\mathrm{np}}=\left(e_{\mathrm{np}x},0,e_{\mathrm{np}z}\right)^{\top}$
and thus the transverse vector has no component in the $y$ direction
$\mathbf{e}_{\mathrm{np}t}=\left(e_{\mathrm{np}x},0\right)^{\top}$.
On the other hand, as in every TM mode, $e_{\mathrm{np}x}$ and $e_{\mathrm{np}z}$
depend on the $x$ coordinate exclusively. The equation we obtained
in the quasi-transverse approximation (\ref{eq:NL_equation_only_transverse})
becomes then a single equation for the $x$ component $e_{\mathrm{np}x}(x)$
in which there is no dependence in the $y$ direction. Remarkably,
there are some cases for which there exists an analytical solution.
That is the case of a planar metal/Kerr structure \cite{Agranovich_JETP_1980}.
The general form of a solution of the stationary transverse problem
(\ref{eq:NL_equation_only_transverse}) for a TM mode with only $x$
component has to be analogous to that of the soliton field in the
previous section:
\[
\bar{E}_{\mathrm{np}x}(x,z)=e_{\mathrm{np}x}(x)e^{i\beta_{\mathrm{np}}z}=Af_{\mathrm{np}}(x;A)e^{i\beta_{\mathrm{np}}z},\,\,\,\,\,\, A\in\mathbb{R}^{+}
\]
where $A$ is the nonlinear plasmon amplitude. As for the soliton
case, we choose it to be the peak value for $e_{\mathrm{np}x}$ ($A=|e_{\mathrm{np}x,0}|$)
whereas $f_{\mathrm{np}}$ plays the role of the \emph{sech }function\emph{.
}Inasmuch the transverse equation only depends on $e_{\mathrm{np}x}$
and not on $e_{\mathrm{np}z}$ we only have a dependence on $A$ and
not on the amplitude of the axial component. In the general case of
a TM mode with coupled axial and transverse components we would have
instead:
\begin{eqnarray*}
e_{\mathrm{np}x}(x) & = & Af_{\mathrm{np}}(x;A,B)\\
e_{\mathrm{np}z}(x) & = & Bg_{\mathrm{np}}(x;A;B).
\end{eqnarray*}

In our case, since we are applying the quasi-transverse approximation,
we keep the dependence on $A$ exclusively. It will be this coefficient
the only one that we will promote to variational parameter. In order
to select our variational \emph{ansatz} we proceed analogously as
for the soliton field in the previous section. We transform the stationary
solution $\bar{E}_{\mathrm{np}x}$ into the variational \emph{ansatz}
$E_{\mathrm{np}x}$ according to the following rule:
\begin{equation}
\bar{E}_{\mathrm{np}x}(x,z)\rightarrow E_{\mathrm{np}x}(x,z)=A(z)f_{\mathrm{np}}(x;|A(z)|)\,\,\,\,\,\, A(z)\in\mathbb{C}.\label{eq:plasmon_ansatz}
\end{equation}
It must be clear now that once the stationary problem has been solved\emph{
}for all values of $A$ (a prerequisite that must be fulfilled\emph{
prior} to the analysis of the variational equations), the function
$f_{\mathrm{np}}$ in Eq.(\ref{eq:plasmon_ansatz}) is perfectly known
for a given value of $A(z)$. In some particular cases, such as in
Ref.\cite{Agranovich_JETP_1980} it is even possible to provide an analytical
expression for the stationary solution and, consequently, also for
$f_{\mathrm{np}}$.

As before, we now introduce the \emph{ansatz }(\ref{eq:plasmon_ansatz})
into the \emph{dynamical} nonlinear plasmon equation for the $x$
component (\ref{eq:NL_plasmon_wave_eq}) to obtain, in the quasi-transverse approximation:
\begin{equation}
\frac{\partial^{2}E_{\mathrm{np}x}}{\partial z^{2}}+\left(\nabla_{t}^{2}+k_{0}^{2}\varepsilon_{p}\right)E_{\mathrm{np}x}-\frac{\partial}{\partial z}\left(\frac{\partial E_{\mathrm{np}z}}{\partial x}\right)-\frac{\partial}{\partial x}\left(\nabla_{t}\cdot\mathbf{E}_{\mathrm{np}t}\right)
\approx
-k_{0}^{2} \gamma\left|\mathbf{E}_{\mathrm{np}t}\right|^{2}E_{\mathrm{np}x}
\label{eq:NL_equation_for_variational_field}
\end{equation}
where $\varepsilon_{L}=\varepsilon_{p}$ is the linear dielectric
function profile of the metal/dielectric interface, as in Eq.(\ref{eq:dielectric_function_MD_interface}).
We take into account now that, according to the variational \emph{ansatz
}(\ref{eq:plasmon_ansatz}), we have (we write $A(z)$ as $|A(z)|\exp i\varphi_{p}(z)$)
\[
\mathbf{E}_{\mathrm{np}t}(x,z)=e^{i\varphi_{p}(z)}\left[\begin{array}{c}
|A(z)|f_{\mathrm{np}}(x;|A(z)|)\\
0
\end{array}\right]=e^{i\varphi_{p}(z)}\left[\begin{array}{c}
e_{\mathrm{np}x}(x;|A(z)|)\\
0
\end{array}\right]=e^{i\varphi_{p}(z)}\mathbf{e}_{\mathrm{np}t}(x;|A(z)|),
\]
where the function $\mathbf{e}_{\mathrm{np}t}(x;|A(z)|)$ is the
solution of the stationary equation (\ref{eq:NL_equation_only_transverse})
with real amplitude $|A(z)|$. Notice that, consequently, $\mathbf{E}_{\mathrm{np}t}(x,z)$ also satisfies the stationary equation (\ref{eq:NL_equation_only_transverse}) with eigenvalue $\beta_{\mathrm{np}}^2 (|A(z)|)$. The function $\mathbf{e}_{\mathrm{np}t}$
verifies the constraint (\ref{eq:contraint_transverse_approx})
in the quasi-transverse approximation. Thus,
due to our variational ansatz, we can also find the corresponding
constraint for for the variational field $\mathbf{E}_{\mathrm{np}t}$:
\begin{equation}
\nabla_{t}\cdot\mathbf{E}_{\mathrm{np}t}=e^{i\varphi_{p}(z)}\nabla_{t}\cdot\mathbf{e}_{\mathrm{np}t}=-i\beta_{\mathrm{np}}E_{\mathrm{np}z}-\mathbf{F}_{\mathrm{np}t}\cdot\mathbf{E}_{\mathrm{np}t}.\label{eq:constraint_Enp}
\end{equation}
Introducing the constraint above in Eq.(\ref{eq:NL_equation_for_variational_field}),
we obtain
\begin{eqnarray}
\frac{\partial^{2}E_{\mathrm{np}x}}{\partial z^{2}}-\frac{\partial}{\partial z}\left(\frac{\partial E_{\mathrm{np}z}}{\partial x}\right)+i\beta_{\mathrm{np}}\frac{\partial E_{\mathrm{np}z}}{\partial x}+\nonumber \\
 &  & \hspace{-3cm}+\left[\nabla_{t}^{2}+k_{0}^{2}\varepsilon_{\mathrm{np}}\right]E_{\mathrm{np}x}+\frac{\partial}{\partial x}\left(\mathbf{F}_{\mathrm{np}t}\cdot\mathbf{E}_{\mathrm{np}t}\right)=0.\label{eq:variational_eq_plasmon}
\end{eqnarray}
%
%
%
%
Taking into account that $\mathbf{E}_{\mathrm{np}t}(x,z)$ verifies the stationary equation (\ref{eq:NL_equation_only_transverse}) with eigenvalue $\beta_{\mathrm{np}}^2 (|A(z)|)$, the equation for the variational field experiments a
notable simplification
\begin{equation}
\frac{\partial^{2}E_{\mathrm{np}x}}{\partial z^{2}}-\frac{\partial}{\partial z}\left(\frac{\partial E_{\mathrm{np}z}}{\partial x}\right)+i\beta_{\mathrm{np}}\frac{\partial E_{\mathrm{np}z}}{\partial x}+\beta_{\mathrm{np}}^{2}E_{\mathrm{np}x}=0.\label{eq:NL_equation_variational_field_plasmon}
\end{equation}
Now we can proceed to project this equation onto a suitable function
in order to find the dynamical equation for the variational parameter
$A(z)$. Since we are dealing with a vector equation for the electromagnetic
field, we should perform the projection using a proper scalar product
for this case. Orthogonality for vector eigenmodes is defined through
the vector relation $\int\left(\mathbf{H}^{*}\times\mathbf{E}\right)\cdot\hat{\mathbf{z}}$
which equals $\int H_{y}E_{x}$ in our case. Thus, we should project
Eq.(\ref{eq:NL_equation_variational_field_plasmon}) onto a suitable
selected value of $H_{y}(x,z)$. As for the soliton case, a natural
choice is to take the stationary solution at the initial propagation
point $\bar{H}_{y}(x,0)$. After performing the projection, we find
\[
\frac{d^{2}}{dz^{2}}\left[\int_{\mathbb{R}}\bar{H}_{y}(0)E_{\mathrm{np}x}(z)\right]-\left(\frac{d}{dz}-i\beta_{\mathrm{np}}\right)\left[\int_{\mathbb{R}}\bar{H}_{y}(0)\left(\frac{\partial E_{\mathrm{np}z}}{\partial x}\right)\right]+\beta_{\mathrm{np}}^{2}\left[\int_{\mathbb{R}}\bar{H}_{y}(0)E_{\mathrm{np}x}(z)\right]=0.
\]
First of all, we pay attention to the integral involving the axial
component:
\begin{eqnarray}
\int_{\mathbb{R}}\bar{H}_{y}(x,0)\frac{\partial}{\partial x}E_{\mathrm{np}z}(x,z) & = & -\int_{\mathbb{R}}\left(\frac{\partial}{\partial x}\bar{H}_{y}(x,0)\right)E_{\mathrm{np}z}(x,z)\nonumber \\
 & = & i\frac{k_{0}}{c}\int_{\mathbb{R}}\varepsilon_{\mathrm{np}}\bar{E}_{\mathrm{np}z}(x,0)E_{\mathrm{np}z}(x,z)\nonumber \\
 & = & O(E_{\mathrm{np}z}^{2})\rightarrow0,\label{eq:integral_axial_component}
\end{eqnarray}
which vanishes in the quasi-transverse approximation. We have used
here the Maxwell's equation $\left[\nabla\times\mathbf{H}\right]_{z}=\partial H_{y}/\partial x=-i\omega c^{-2}\varepsilon E_{z}$
to write the previous integral in terms of the axial component of
the electric field.

Now we consider the realistic situation in which the system admit
losses, a circumstance which is immediate in the case of metals. We
return to the dynamical equation for the variational field (\ref{eq:variational_eq_plasmon})
and consider now that $\varepsilon_{p}$ is a complex function, so
that, we make the substitution $\varepsilon_{p}\rightarrow\varepsilon_{p}+i\triangle_{l}\varepsilon_{p}$
, where we keep in our notation $\varepsilon_{p}$ as the real part
of the dielectric function whereas $\triangle_{l}\varepsilon_{p}$
is a function indicating the distribution of linear losses in the
system. This substitution generates an extra term in Eqs.(\ref{eq:variational_eq_plasmon})
and (\ref{eq:NL_equation_variational_field_plasmon}) of the form
$ik_{0}^{2}\triangle_{l}\varepsilon_{p}E_{\mathrm{np}x}$, in such
a way that, after the projection, we obtain:
\[
\frac{d^{2}}{dz^{2}}\left[\int_{\mathbb{R}}\bar{H}_{y}(0)E_{\mathrm{np}x}(z)\right]+\beta_{\mathrm{np}}^{2}\left[\int_{\mathbb{R}}\bar{H}_{y}(0)E_{\mathrm{np}x}(z)\right]+ik_{0}^{2}\left[\int_{\mathbb{R}}\bar{H}_{y}(0)\triangle_{l}\varepsilon_{p}E_{\mathrm{np}x}(z)\right]=0.
\]
We recall that for a TM stationary mode it is true that
\[
\bar{H}_{y}(x,0)=\frac{k_{0}c}{\beta_{\mathrm{np}}(0)}\bar{E}_{\mathrm{np}x}(x,0)=Kf_{\mathrm{np}}(x;|A(0)|),
\]
where $K$ is a constant independent of $x$ and $z$, so that it
disappears from the equation. On the other hand, 
\[
E_{\mathrm{np}x}(x,z)=A(z)f_{\mathrm{np}}(x;|A(z)|).
\]
Therefore,
\begin{equation}
\frac{d^{2}}{dz^{2}}\left[N_{\mathrm{np}}A(z)\right]+N_{\mathrm{np}}\left[\beta_{\mathrm{np}}^{2}+i\lambda_{\mathrm{np}}\right]A(z)=0,\label{eq:variational_eq_for_A}
\end{equation}
where we have defined the ``norm'' $N_{\mathrm{np}}$ as
\begin{equation}
N_{\mathrm{np}}\equiv\int_{\mathbb{R}}f_{\mathrm{np}}(x;|A(0)|)\mathrm{f}_{\mathrm{np}}(x;|A(z)|)\label{eq:norm_plasmon}
\end{equation}
and the loss parameter $\lambda_{\mathrm{np}}$ as
\[
\lambda_{\mathrm{np}}\equiv\frac{1}{N_{\mathrm{np}}}\int_{\mathbb{R}}f_{np}(x;|A(0)|)\triangle_{l}\varepsilon_{p}(x)f_{\mathrm{np}}(x;|A(z)|).
\]
By construction, the ``norm'' $N_{\mathrm{np}}$ and the loss parameter
$\lambda_{\mathrm{np}}$ depend on the modulus of the variational
parameter $|A(z)|$. Likewise, the propagation constant $\beta_{\mathrm{np}}$
of the nonlinear plasmon is also dependent on this quantity. All these
dependence on $|A(z)|$ can be fully established once the stationary
nonlinear problem has been thoroughly solved. A further simplification
of the variational equation is permitted in the quasi-stationary approximation
$d|A|/dz\ll d\varphi_{p}/dz$, since we can proceed as we did for
the soliton case to write 
\[
\frac{d^{2}}{dz^{2}}\left[N_{\mathrm{np}}\left(|A(z)|\right)A(z)\right]\approx N_{\mathrm{np}}\left(|A(z)|\right)\left(\frac{d^{2}A(z)}{dz^{2}}\right),
\]
so that we finally obtain
\[
N_{\mathrm{np}}\left[\frac{d^{2}}{dz^{2}}A(z)+\bar{\beta}_{\mathrm{np}}^{2}\left(|A|\right)A(z)\right]=0,
\]
where we have defined a complex effective nonlinear propagation constant
as $\bar{\beta}_{\mathrm{np}}^{2}\equiv\beta_{\mathrm{np}}^{2}+i\lambda_{\mathrm{np}}$.

\section{Variational equations for the soliplasmon bound state\label{sec:variational_eq_for_soliplasmon}}

In this section we will establish the variational equations describing
the propagation of a soliplasmon bound state. Our variational \emph{ansatz
}for a soliplasmon will be given by a superposition of a nonlinear
plasmon and a soliton. However, instead of assuming a general dependence
in multiple variational parameters, as in Eq.(\ref{eq:general_ansatz}),
we will reduce the number of variational parameters just to two: one
associated to the nonlinear plasmon, $A(z)$, and a second one, associated
to the soliton, $C(z)$. They correspond to the amplitudes of the
nonlinear plasmon and soliton solutions as defined in Eqs.(\ref{eq:soliton_ansatz})
and (\ref{eq:plasmon_ansatz}). Therefore, our variational \emph{ansatz}
for the soliplasmon solution will be: 

\emph{
\begin{equation}
\mathbf{E}(x,z)=\mathbf{E}_{\mathrm{np}}\left(x;A\left(z\right)\right)+\mathbf{u}E_{s}\left(x;C\left(z\right)\right).\label{eq:soliplasmon_ansatz}
\end{equation}
}

In the same way, we will treat mathematically the plasmon and soliton
components as we did in the previous two sections. This means that
we will assume the same approximations we used to demonstrate the
variational equations for the uncoupled system. Summarizing, we will
work under the following approximations for the variational fields:
\begin{itemize}
\item \emph{Scalar approximation} for the soliton field: $\nabla\cdot\mathbf{E}_{s}\approx0$.
\item \emph{Quasi-transverse approximation} for the plasmon field. Terms
of order $ $$E_{\mathrm{np}z}^{2}$ and higher will be neglected:
$O(E_{\mathrm{np}z}^{2})\rightarrow0$.
\item \emph{Quasi-stationary approximation} for both: $d|C|/dz\ll d\varphi_{s}/dz$
and $d|A|/dz\ll d\varphi_{p}/dz$.
\end{itemize}

\subsection{Equations for the coupled plasmon and soliton variational fields}

Taking all these approximations into account we proceed to find the
corresponding variational equations for the plasmon and soliton field
components of our \emph{ansatz} (\ref{eq:soliplasmon_ansatz}). As
mentioned in Section \ref{sec:variational_ansatz_NL_maxwell_eqs},
when the soliton is located infinitely far away from the nonlinear
plasmon, the two localized solutions in Eq.(\ref{eq:soliplasmon_ansatz})
present a vanishing overlapping, so they can be treated independently
in such a way they verify uncoupled independent equations. This is
the analysis we have performed in the two previous sections. When
this overlap cannot be neglected, an explicit coupling appears and
then Eq.(\ref{eq:wave_eq_Enp+Es}) holds instead. On the other hand,
the soliplasmon solution found in Ref.\cite{Milian_OL_2012} is a TM
mode of the electromagnetic field. Besides, the axial component of
its electric field is only relevant for the plasmon field close to
the interface and not for the soliton. For these reasons, we consider
that $\mathbf{E}=(E_{x},0,E_{z})$ and $\mathbf{u}\approx(1,0,0)$.$ $$ $
In this way, the axial component will be approximately given by the
plasmonic component exclusively ($E_{z}\approx E_{\mathrm{np}z}$)
so that it will be possible to evaluate it from $E_{\mathrm{np}x}$
according to the procedure presented in the previous section. Consequently,
our starting point will be the variational equation for the $x$ component
of the electric field. So that, we write Eq.(\ref{eq:wave_eq_Enp+Es})
for the $x$ component (recall that $L_{0}\equiv\nabla_{t}^{2}+k_{0}^{2}\varepsilon_{L}$)
\begin{eqnarray}
\frac{\partial^{2}E_{\mathrm{np}x}}{\partial z^{2}}-\frac{\partial}{\partial z}\left(\frac{\partial E_{\mathrm{np}z}}{\partial x}\right)+i\beta_{\mathrm{np}}\frac{\partial E_{\mathrm{np}z}}{\partial x}+L_{0}E_{\mathrm{np}x}+\frac{\partial}{\partial x}\left(\mathbf{F}_{\mathrm{np}t}\cdot\mathbf{E}_{\mathrm{np}t}\right)+\frac{\partial^{2}E_{s}}{\partial z^{2}}+L_{0}E_{s} & =\nonumber \\
 &  & \hspace{-10cm}-k_{0}^{2}\gamma\left|\mathbf{E}_{\mathrm{np}t}\right|^{2}E_{\mathrm{np}x}-k_{0}^{2}\gamma\left|E_{s}\right|^{2}E_{s}-k_{0}^{2}Q_{\mathrm{K}}(E_{\mathrm{npx}},E_{s}),\label{eq:wave_eq_Enp+Es_x_comp}
\end{eqnarray}
where we have used the constraint (\ref{eq:constraint_Enp}) for $\nabla\cdot\mathbf{E}_{\mathrm{np}}$
---valid in the quasi-transverse approximation--- and defined $Q_{\mathrm{K}}\equiv\mathbf{Q}_{\mathrm{K},x}$.

There are two coupling mechanisms in Eq.(\ref{eq:wave_eq_Enp+Es_x_comp}).
One is purely nonlinear and it is generated by the Kerr coupling term
$Q_{\mathrm{K}}$. The other one is due to the presence of a variation
of the linear dielectric function in the regions where the field is
localized, i.e, in the nonlinear plasmon and in the soliton regions,
in our case. This is a well known mechanism in solid state physics
and it is the origin of the coupling between neighboring wave functions
in the so-called \emph{tight binding approximation }\cite{Ashcroft_book_1976}\emph{.
}Let us see how it works in the present case. We note that the linear
operator $L_{0}$ in Eq.(\ref{eq:wave_eq_Enp+Es_x_comp}) does not
coincides exactly with that corresponding to the uncoupled solutions
in the variational equations for the soliton (Eq.(\ref{eq:variational_eq_soliton}))
and nonlinear plasmon (Eq.(\ref{eq:variational_eq_plasmon})). The
reason is that the total \emph{linear} dielectric function differs
from the uncoupled ones in the regions of the 1D space where the functions
are not localized. To be more specific, let us write this function
for a MDK structure as in Fig.\ref{fig:MDK}:
\begin{equation}
\varepsilon_{L}(x)=\begin{cases}
\varepsilon_{p}(x) & x\le d\\
\varepsilon_{\mathrm{K}} & x>d
\end{cases},\label{eq:dielectric_function_MDK}
\end{equation}
where $d$ is the thickness of the dielectric layer. In this case,
$\varepsilon_{p}$ represents the profile of the \emph{linear} dielectric
function for the plasmon component, i.e., it defines the dielectric
constant profile of the MD structure, as defined in Eq.(\ref{eq:dielectric_function_MD_interface}).
We could consider, with any lack of generality, that there is also
a modulation of the \emph{linear} dielectric function in the Kerr
medium, so that, we would have a function $\varepsilon_{s}(x)$ instead
of $\varepsilon_{\mathrm{K}}$ in the previous expression. However,
in order to preserve the analysis of the soliton component exactly
as we did in Section \ref{sec:variational_eq_soliton} we will keep
$\varepsilon_{\mathrm{K}}$ as the linear dielectric function for
the soliton field. Generalization to arbitrary $\varepsilon_{s}(x)$
will be straightforward once final results are obtained.

The definition of the total dielectric function in Eq.(\ref{eq:dielectric_function_MDK})
suggests the following two decompositions for $\varepsilon_{L}$:
\begin{eqnarray}
\varepsilon_{L} & = & \varepsilon_{p}+\triangle\varepsilon_{p}\nonumber \\
\varepsilon_{L} & = & \varepsilon_{\mathrm{K}}+\triangle\varepsilon_{\mathrm{s}},\label{eq:decomposition_linear_dielectric_function}
\end{eqnarray}
 where the local variations $\triangle\varepsilon_{p}$ and $\triangle\varepsilon_{s}$
would be given by
\begin{equation}
\triangle\varepsilon_{p}(x)\,\,\,=\,\,\,\begin{cases}
0 & x\le d\\
\varepsilon_{\mathrm{K}}-\epsilon_{p}(x) & x>d
\end{cases}\,\,\,=\,\,\,\begin{cases}
0 & x\le d\\
\varepsilon_{\mathrm{K}}-\varepsilon_{d} & x>d
\end{cases}\label{eq:local_variation_dielectric_function_plasmon}
\end{equation}
 and
\begin{equation}
\triangle\varepsilon_{s}(x)\,\,\,=\,\,\,\begin{cases}
\varepsilon_{p}(x)-\epsilon_{\mathrm{K}} & x\le d\\
0 & x>d
\end{cases}\,\,\,=\,\,\,\begin{cases}
\varepsilon_{m}-\varepsilon_{\mathrm{K}} & x\leq0\\
\varepsilon_{d}-\varepsilon_{\mathrm{K}} & 0<x\le d\\
0 & x>d
\end{cases},\label{eq:local_variation_dielectric_function_soliton}
\end{equation}
where we have taken into account the dielectric function profile for
the MD interface as given by Eq.(\ref{eq:dielectric_function_MD_interface}).
The decomposition (\ref{eq:decomposition_linear_dielectric_function})
of the linear dielectric function permits to write the operator $L_{0}$
in two different ways:
\begin{eqnarray*}
L_{0} & = & \left(\nabla_{t}^{2}+k_{0}^{2}\varepsilon_{p}\right)+\triangle\varepsilon_{p}\equiv L_{0p}+\triangle\varepsilon_{p}\\
L_{0} & = & \left(\nabla_{t}^{2}+k_{0}^{2}\varepsilon_{\mathrm{K}}\right)+\triangle\varepsilon_{s}\equiv L_{0s}+\triangle\varepsilon_{s}
\end{eqnarray*}
 and, therefore, we can rewrite Eq.(\ref{eq:wave_eq_Enp+Es_x_comp})
as
\begin{eqnarray*}
\frac{\partial^{2}E_{\mathrm{np}x}}{\partial z^{2}}-\frac{\partial}{\partial z}\left(\frac{\partial E_{\mathrm{np}z}}{\partial x}\right)+i\beta_{\mathrm{np}}\frac{\partial E_{\mathrm{np}z}}{\partial x}+\underbrace{\left[L_{0p}+k_{0}^{2}\gamma\left|\mathbf{E}_{\mathrm{np}t}\right|^{2}\right]E_{\mathrm{np}x}+\frac{\partial}{\partial x}\left(\mathbf{F}_{\mathrm{np}t}\cdot\mathbf{E}_{\mathrm{np}t}\right)}_{\mathrm{eigevalue\,\, equation\,\, for\,\, NL\,\, plasmon}} & +\\
+\frac{\partial^{2}E_{s}}{\partial z^{2}}+\underbrace{\left[L_{0s}+k_{0}^{2}\gamma\left|E_{s}\right|^{2}\right]E_{s}}_{\mathrm{eigevalue\,\, equation\,\, for\,\, soliton}}=-k_{0}^{2}\left(\triangle\varepsilon_{p}E_{\mathrm{np}x}+\triangle\varepsilon_{s}E_{s}\right)-k_{0}^{2}Q_{\mathrm{K}}(E_{\mathrm{npx}},E_{s}).
\end{eqnarray*}
We immediately recognize in the previous expression the appearance
of the nonlinear operators for the soliton and plasmon stationary
solutions. We demonstrated that soliton and plasmon variational fields
were also eigefunctions of these operators,
so that a simplification
of the above equation can be obtained:
\begin{eqnarray}
\frac{\partial^{2}E_{\mathrm{np}x}}{\partial z^{2}}-\left(\frac{\partial}{\partial z}-i\beta_{\mathrm{np}}\right)\frac{\partial E_{\mathrm{np}z}}{\partial x}+\beta_{\mathrm{np}}^{2}E_{\mathrm{np}x}+\frac{\partial^{2}E_{s}}{\partial z^{2}}+\beta_{s}^{2}E_{s} & =\nonumber \\
 &  & \hspace{-5cm}-k_{0}^{2}\left(\triangle\varepsilon_{p}E_{\mathrm{np}x}+\triangle\varepsilon_{s}E_{s}\right)-k_{0}^{2}Q_{\mathrm{K}}(E_{\mathrm{npx}},E_{s}).\label{eq:NL_equation_variational_field_soliplasmon}
\end{eqnarray}
\subsection{Dynamical equations for the variational parameters in the weak coupling
approximation}

In order to obtain the equations for the variational parameter $A(z)$
and $C(z)$ we need to project out Eq.(\ref{eq:NL_equation_variational_field_soliplasmon})
into the adequate projection functions. In the two previous sections
we projected the two uncoupled equations for the soliton and plasmon
variational fields using suitable soliton and plasmon projection functions
for each case. Here, we will make an identical choice for the projecting
functions and we will make two different projections of Eq.(\ref{eq:NL_equation_variational_field_soliplasmon}).
The first one, corresponding to the soliton projection, will be performed
with respect to the soliton stationary field at $z=0$, i.e., $\bar{E}_{s}(x,0)$.
For the second one, corresponding to the plasmonic projection, we
will use the plasmon stationary field at $z=0$, i.e., $\bar{H}_{y}(x,0)$.

When performing the aforementioned projections in Eq.(\ref{eq:NL_equation_variational_field_soliplasmon}),
we will encountered a new type of overlapping integrals not present
in our previous analysis. They correspond to integrals involving products
of soliton and plasmon fields. Since soliton and plasmon functions
are localized in different regions of the space, these integrals are
expected to be small when the soliton field is localized sufficiently
far way from the MD interface. More specifically, if we analyze an
overlapping integral of the form (we assume $F_{p}$ to be a plasmonic
function tightly localized around the interface)
\[
I_{n}(a)=\int_{\mathbb{R}}F_{p}(x)f_{s}^{n}(x-a),
\]
and we consider the overlapping to be small ($a\gg1$, implying $|I_{n}|\ll1$),
we can estimate the order of this integral by approximating the \emph{sech}
function by its exponential tail, so that, $f_{s}\approx\exp\kappa_{s}\left(x-a)\right)$,
where $\kappa_{s}=\left(\gamma/2\right)^{1/2}k_{0}\left|C\right|$. Therefore,
\begin{equation}
I_{n}(a)\approx e^{-n\kappa_{s}a}\int_{0}^{d_{p}}F_{p}(x)e^{n\kappa_{s}x}\sim O(e^{-\kappa_{s}a})^{n},\label{eq:integral_plasmon_soliton_small}
\end{equation}
where $d_{p}$ is a small parameter of the order of the penetration
length of the localized function $F_{p}$ into the dielectric medium.
The previous argument suggests to take $\epsilon_{s}=e^{-\kappa_{s}a}$
as a small parameter. Analogously, terms depending on the plasmonic
tail of the form $e^{-\kappa_{p}d}$ where the distance $d$ is substantially
larger than the plasmon penetration length ($d\gg\kappa_{p}^{-1}$)
will be also small, which defines $\epsilon_{p}=e^{-\kappa_{p}d}$
as a small parameter as well. Hence, we define the following additional
approximation associated to the coupling of soliton and plasmonic
components:
\begin{itemize}
\item \emph{Weak coupling approximation}.\label{weak-coupling-approximation} 
\begin{itemize}
\item Terms of order $ $$e^{-2\kappa_{s}a}$ and higher will be neglected:
$O(e^{-2\kappa_{s}a})=O(\epsilon_{s}^{2})\rightarrow0$. 
\item Terms of order $e^{-2\kappa_{p}d}$ or higher will be likewise neglected:
$O(e^{-2\kappa_{p}d})=O(\epsilon_{p}^{2})\rightarrow0$.
\end{itemize}
\end{itemize}
The previous approximation complete the set stated previously. Now,
we use it to perform the projections by keeping only the leading terms. 

We can further simplify the soliplasmon equation for the variational
fields (\ref{eq:NL_equation_variational_field_soliplasmon}) by simultaneously
invoking the quasi-transverse and weak coupling approximations. Let
us pay attention to the second term depending on the axial component
$E_{\mathrm{np}z}$. The plasmonic projection of this term provides
the overlapping integral already seen in Eq.(\ref{eq:integral_axial_component}),
which, it was proven to be $O(E_{\mathrm{np}z}^{2})$ and, thus, negligible.
The soliton projection provides, in turn, the integral
\[
\int_{\mathbb{R}}E_{\mathrm{np}z}(x)f_{s}(x-a)\approx e^{-\kappa_{s}a}\int_{0}^{d_{p}}E_{\mathrm{np}z}(x)e^{n\kappa_{s}x}\sim O(e^{-\kappa_{s}a})O(E_{\mathrm{np}z})\rightarrow0,
\]
which we also neglect as a product of two infinitesimals. Therefore,
this term does not contribute to any of the two projections and it
can be also neglected. 

However, this is not the only simplification that we can perform using
the weak coupling approximation. They also apply to the nonlinear
coupling term $Q_{\mathrm{K}}(E_{\mathrm{npx}},E_{s})$. From its
definition (\ref{eq:NL_Kerr_coupling}) we have
\begin{equation}
Q_{\mathrm{K}}(E_{\mathrm{npx}},E_{s})=\gamma\left[2|E_{s}|^{2}E_{\mathrm{np}x}+E_{s}^{2}E_{\mathrm{np}x}^{*}+2|E_{\mathrm{np}x}|^{2}E_{s}+E_{\mathrm{np}x}^{2}E_{s}^{*}\right].\label{eq:general_Kerr_coupling}
\end{equation}
We immediately see from Eq.(\ref{eq:integral_plasmon_soliton_small})
that both the plasmonic and soliton projections of the two quadratic
terms in $E_{s}$ of $Q_{\mathrm{K}}$ are, at least, $O(e^{-2\kappa_{s}a})$ 
\footnote{Recall that $E_{s}=Cf_{s}$ and $\bar{H}_{y}=k_{0}c\beta_{\mathrm{np}}^{-1}\bar{E}_{\mathrm{np}x}$,
so that the plasmonic projection gives
\[
\int_{\mathbb{R}}\bar{H}_{y}(0)|E_{s}|^{2}E_{\mathrm{np}x}\sim\int\bar{E}_{\mathrm{np}x}(0)E_{\mathrm{np}x}f_{s}^{2}\sim O(e^{-2\kappa_{s}a}),
\]
whereas the soliton one yields
\[
\int_{\mathbb{R}}\bar{f}_{s}(0)E_{\mathrm{np}x}f_{s}^{2}\sim O(e^{-\left(2\kappa_{s}+\kappa_{s}(0)\right)a}).
\]
Both are negligible in the weak coupling approximation. Analogous
argument holds for the $E_{s}^{2}E_{\mathrm{np}x}^{*}$ term.%
}. Therefore, the first two terms in the previous equation can also
be neglected. 

Taking into account all the previous approximations, our final equation
for the variational fields takes a relatively simple form:
\begin{eqnarray}
\frac{\partial^{2}E_{\mathrm{np}x}}{\partial z^{2}}+\beta_{\mathrm{np}}^{2}E_{\mathrm{np}x}+\frac{\partial^{2}E_{s}}{\partial z^{2}}+\beta_{s}^{2}E_{s} & \approx\nonumber \\
 &  & \hspace{-5cm}-k_{0}^{2}\left(\triangle\varepsilon_{p}E_{\mathrm{np}x}+\triangle\varepsilon_{s}E_{s}\right)-k_{0}^{2}\gamma\left[2|E_{\mathrm{np}x}|^{2}E_{s}+E_{\mathrm{np}x}^{2}E_{s}^{*}\right].\label{eq:NL_equation_variational_field_soliplasmon_simplified}
\end{eqnarray}
Our final step is to project Eq.(\ref{eq:NL_equation_variational_field_soliplasmon_simplified})
into the soliton and plasmon projection functions in order to convert
this equation in two different equations for $A(z)$ and $C(z)$.

\subsubsection{Plasmon projection}

We start now by performing the plasmonic projection first. As in Section
\ref{sec:variational_equation_for_NL_plasmon}, we project with respect
to $\bar{H}_{y}(x,0)=Kf_{\mathrm{np}}(x;|A(0)|),$ where the constant
$K$ can be ignored since it disappears after the projection. We obtain
\begin{equation}
N_{\mathrm{np}}\left[\frac{d^{2}}{dz^{2}}A+\beta_{\mathrm{np}}^{2}A\right]+\delta_{ps}\left[\frac{d^{2}}{dz^{2}}C+\beta_{\mathrm{s}}^{2}C\right]=-\triangle_{pp}A-\triangle_{ps}C-\triangle_{\mathrm{K}}\left[2|A|^{2}C+A^{2}C^{*}\right],\label{eq:plasmonic_projection}
\end{equation}
where $N_{\mathrm{np}}$ is defined as in Eq.(\ref{eq:norm_plasmon})
and we have defined three new coupling terms
\begin{eqnarray}
\delta_{ps} & \equiv & \int_{\mathbb{R}}f_{\mathrm{np}}(0)f_{s}(z)\nonumber \\
\triangle_{ps} & \equiv & k_{0}^{2}\int_{\mathbb{R}}f_{\mathrm{np}}(0)\triangle\varepsilon_{s}f_{s}(z)\nonumber \\
\triangle_{\mathrm{K}} & \equiv & k_{0}^{2}\int_{\mathbb{R}}\gamma f_{\mathrm{np}}(0)f_{\mathrm{np}}^{2}(z)f_{s}(z).\label{eq:plasmon_projection_integrals}
\end{eqnarray}
However, the plasmonic self-interaction term $\triangle_{pp}$ term
can be neglected according to the weak coupling approximation since
\begin{equation}
\triangle_{pp}\equiv k_{0}^{2}\int_{\mathbb{R}}f_{\mathrm{np}}(0)\triangle\varepsilon_{p}f_{\mathrm{np}}(z)\sim k_{0}^{2}\left(\varepsilon_{\mathrm{k}}-\varepsilon_{d}\right)\int_{d}^{\infty}e^{-2\kappa_{p}x}\sim O(e^{-2\kappa_{p}d})\rightarrow0,\label{eq:Delta_pp_coefficient}
\end{equation}
where we have taken into account the form of the local variation of
the dielectric function for the plasmon as in Eq.(\ref{eq:local_variation_dielectric_function_plasmon}) and we assume that the width of the dielectric layer $d$ is larger than the plasmon penetration length in the dielectric: $d \gg \kappa_d^{-1}$. 
Besides, we have approximated $f_{\mathrm{np}}$ by its linear counterpart
$f_{\mathrm{np}}\approx f_{p}\sim e^{-\kappa_{p}x}$ because in the
regions where de plasmon field is very weak nonlinear effects are
negligible.

In order to obtain the plasmonic projection (\ref{eq:plasmonic_projection})
we have also used the quasi-stationary approximation for the plasmon
field $d|A|/dz\ll d\varphi_{p}/dz$ that we already used in Section
\ref{sec:variational_equation_for_NL_plasmon} to write $d^{2}(N_{\mathrm{np}}A)/dz^{2}\approx N_{\mathrm{np}}d^{2}A/dz^{2}$.
An analogous argument, in this case using the quasi-stationary approximation
for the soliton field $d|C|/dz\ll d\varphi_{s}/dz$, has been also
used to prove that $ $ $d^{2}(\delta_{ps}C)/dz^{2}\approx\delta_{ps}d^{2}C/dz^{2}$.

\subsubsection{Soliton projection}

In this case, we project Eq.(\ref{eq:NL_equation_variational_field_soliplasmon_simplified})
with respect to the stationary soliton field at $z=0$. This field
is given by the same projection function, $\bar{E}_{s}(0)=C(0)f_{s}(0))$,
that we used in Section \ref{sec:variational_eq_soliton}. Since $C(0)$
is a constant not depending on $x$ nor $z$, it disappears after
the projection is carried out, so only the spatial function $f_{s}(x;|C(0)|)$
appears in the projection integrals. In this way, the resulting equation
obtained after soliton projection is:
\begin{equation}
\delta_{sp}\left[\frac{d^{2}}{dz^{2}}A+\beta_{\mathrm{np}}^{2}A\right]+N_{s}\left[\frac{d^{2}}{dz^{2}}C+\beta_{\mathrm{s}}^{2}C\right]=-\triangle_{ss}C-\triangle_{sp}A-\triangle'_{\mathrm{K}}\left[2|A|^{2}C+A^{2}C^{*}\right],\label{eq:soliton_projection}
\end{equation}
where slightly different coupling terms from those appearing in the
plasmonic projection ---Eqs.(\ref{eq:plasmon_projection_integrals})---
are obtained:
\begin{eqnarray}
\delta_{sp} & \equiv & \int_{\mathbb{R}}f_{s}(0)f_{\mathrm{np}}(z)\nonumber \\
\triangle_{sp} & \equiv & k_{0}^{2}\int_{\mathbb{R}}f_{s}(0)\triangle\varepsilon_{p}f_{\mathrm{np}}(z)\nonumber \\
\triangle'_{\mathrm{K}} & \equiv & k_{0}^{2}\int_{\mathbb{R}}\gamma f_{s}(0)f_{\mathrm{np}}^{2}(z)f_{s}(z).
\label{eq:soliton_projection_integral}
\end{eqnarray}
The linear self-coupling term $\triangle_{ss}$ can be approximated
in the weak coupling approximation as follows
\[
\triangle_{ss}\equiv k_{0}^{2}\int_{\mathbb{R}}f_{s}(0)\triangle\varepsilon_{s}f_{\mathrm{s}}(z)\sim k_{0}^{2}\int_{-\infty}^{d}\left[\varepsilon_{p}(x)-\epsilon_{\mathrm{K}}\right]e^{\left[\kappa_{s}(0)+\kappa_{s}\right]\left(x-a\right)}\sim O(e^{-\left[\kappa_{s}(0)+\kappa_{s}\right]a})\rightarrow0,
\]
and, therefore, neglected as its plasmonic counterpart. In the previous
integral we have approximated the \emph{sech} function by its exponential
tail. This is justified because the integral only covers the metal
and dielectric part, so that if the soliton field is located not too
close to the dielectric region, its value in the integral domain will
be given by its exponential tail. Using identical argument, it is
proven that
\[
\triangle'_{\mathrm{K}}\sim O(e^{-\left[\kappa_{s}(0)+\kappa_{s}\right]a})\rightarrow0.
\]
Obviously, it is implicitly assumed that the varying value of $\kappa_{s}(z)$
always satisfies the weak coupling condition. So, if the condition
is satisfied by the field at $z=0$ ---determined by the value of
$\kappa_{s}(0)$---, then $ $$\kappa_{s}(z)\sim\kappa_{s}(0)$ for
all values of $z$ so that:
\[
O(e^{-\kappa_{s}(z)a})\sim O(e^{-\kappa_{s}(0)a})\,\,\,\,\,\,\forall z,
\]
and, therefore,
\[
O(e^{-\left[\kappa_{s}(0)+\kappa_{s}\right]a})\sim O(e^{-2\kappa_{s}(0)a})\sim O(e^{-2\kappa_{s}a})\rightarrow0.
\]
Consequently, the weak coupling condition for the soliton field in
Section \ref{weak-coupling-approximation} should be understood in
the above sense.

As for the plasmonic projection, we have also used the quasi-stationary
approximation to write $d^{2}(\delta_{sp}A)/dz^{2}\approx\delta_{sp}d^{2}A/dz^{2}$
and $d^{2}(N_{s}C)/dz^{2}\approx N_{s}d^{2}C/dz^{2}$.

\subsubsection{Equations for the variational parameters $A(z)$ and $C(z)$}

The plasmonic and soliton projections above provide us already with
dynamical equations for the variational parameters, but they are not
yet in their ``canonical form.'' In order to see this feature, we
write them again together using matrix notation
\begin{eqnarray*}
\left[\begin{array}{cc}
N_{\mathrm{np}} & \delta_{ps}\\
\delta_{sp} & N_{s}
\end{array}\right]\left[\begin{array}{c}
\left(\frac{d^{2}}{dz^{2}}+\beta_{\mathrm{np}}^{2}\right)A\\
\left(\frac{d^{2}}{dz^{2}}+\beta_{s}^{2}\right)C
\end{array}\right] & = & -\left[\begin{array}{cc}
\triangle_{pp} & \triangle_{ps}\\
\triangle_{sp} & \triangle_{ss}
\end{array}\right]\left[\begin{array}{c}
A\\
C
\end{array}\right]-\left[\begin{array}{cc}
\triangle_{\mathrm{K}}AC^{*} & 2\triangle_{\mathrm{K}}|A|^{2}\\
\triangle'_{\mathrm{K}}AC^{*} & 2\triangle'_{\mathrm{K}}|A|^{2}
\end{array}\right]\left[\begin{array}{c}
A\\
C
\end{array}\right].
\end{eqnarray*}
Although we know that some of the coefficients in the previous equation
vanish in the weak coupling approximation, as we have just proven,
we will keep them in order to analyze some interesting particular
case in the next section. So, by pre-multiplying the equation above
by
\[
\left[\begin{array}{cc}
N_{\mathrm{np}} & \delta_{ps}\\
\delta_{sp} & N_{s}
\end{array}\right]^{-1}
\]
we can set the variational equations in their standard form:
\begin{eqnarray}
\left(\frac{d^{2}}{dz^{2}}+B_{\mathrm{np}}^{2}\right)A & =- & q_{ps}C-q_{\mathrm{K}}\left(2|A|^{2}C+A^{2}C^{*}\right)\nonumber \\
\left(\frac{d^{2}}{dz^{2}}+B_{s}^{2}\right)C & =- & q_{sp}A-q'_{\mathrm{K}}\left(2|A|^{2}C+A^{2}C^{*}\right).\label{eq:variational_eq_soliplasmon_parameters}
\end{eqnarray}
We can recognize that there exist three type of terms in the previous
variational equations:
\begin{itemize}
\item Terms modifying the phase velocity of the plasmon and soliton components
by renormalizing their propagation constant through terms lineal in
$A$ and $C$, respectively:
\begin{eqnarray*}
B_{\mathrm{np}}^{2} & \equiv & \beta_{\mathrm{np}}^{2}-\left(\frac{\text{\ensuremath{\delta}}_{ps}\text{\ensuremath{\Delta}}_{ps}-N_{s}\text{\ensuremath{\Delta}}_{pp}}{N_{\mathrm{np}}N_{s}-\text{\ensuremath{\delta}}_{ps}\ensuremath{\delta_{sp}}}\right)\\
B_{s}^{2} & \equiv & \beta_{s}^{2}-\left(\frac{\ensuremath{\delta}_{sp}\text{\ensuremath{\Delta}}_{sp}-N_{\mathrm{np}}\Delta_{ss}}{N_{\mathrm{np}}N_{s}-\text{\ensuremath{\delta}}_{ps}\ensuremath{\delta_{sp}}}\right).
\end{eqnarray*}

\item Terms coupling plasmon and soliton components, which are \emph{linear}
in $A$ and $C$, respectively:
\begin{eqnarray}
q_{ps} & \equiv & \left(\frac{N_{s}\text{\ensuremath{\Delta}}_{ps}-\text{\ensuremath{\delta}}_{ps}\text{\ensuremath{\Delta}}_{ss}}{N_{\mathrm{np}}N_{s}-\text{\ensuremath{\delta}}_{ps}\ensuremath{\delta_{sp}}}\right)\nonumber \\
q_{sp} & \equiv & \left(\frac{N_{\mathrm{np}}\ensuremath{\Delta}_{sp}-\text{\ensuremath{\delta}}_{sp}\text{\ensuremath{\Delta}}_{pp}}{N_{\mathrm{np}}N_{s}-\text{\ensuremath{\delta}}_{ps}\ensuremath{\delta_{sp}}}\right).\label{eq:q_linear_coefficients}
\end{eqnarray}

\item Terms coupling plasmon and soliton components, which are \emph{nonlinear}
in $A$ and $C$, respectively:
\begin{eqnarray*}
q_{\mathrm{K}} & \equiv & \left(\frac{N_{s}\Delta_{\mathrm{K}}-\delta_{ps}\Delta'_{\mathrm{K}}}{N_{\mathrm{np}}N_{s}-\text{\ensuremath{\delta}}_{ps}\ensuremath{\delta_{sp}}}\right)\\
q'_{\mathrm{K}} & \equiv & \left(\frac{N_{\mathrm{np}}\ensuremath{\Delta}'_{\mathrm{K}}-\text{\ensuremath{\delta}}_{sp}\text{\ensuremath{\Delta}}_{\mathrm{K}}}{N_{\mathrm{np}}N_{s}-\text{\ensuremath{\delta}}_{ps}\ensuremath{\delta_{sp}}}\right).
\end{eqnarray*}

\end{itemize}
Thus the evolution equations for the variational parameters of the
soliplasmon problem (\ref{eq:variational_eq_soliplasmon_parameters}) can be written as a system of coupled nonlinear
resonators with special characteristics. The special features we are referring to have to do with the nonlinear
dependence of all the coefficients in the previous equations on the
modulus of the variational parameters $|A|$ and $|C|$. All coefficients
depend on quantities which are given in terms of integrals of the
stationary functions $f_{\mathrm{np}}(x;|A(z)|)$ and $f_{s}(x;|C(z)|)$.
Note that these nonlinear terms are different from those owned
by the nonlinear plasmon and soliton when they are decoupled. In the absence of coupling, the nonlinearities come from the dependence
of $\beta_{\mathrm{np}}$ and $\beta_{s}$ on $|A|$ and $|C|$, respectively
(as we have seen in Sections \ref{sec:variational_eq_soliton} and
\ref{sec:variational_equation_for_NL_plasmon}.) Note as well that
there are nonlinear dependences which are not explicitly given by
the obvious crossed Kerr terms in Eqs.(\ref{eq:variational_eq_soliplasmon_parameters}).
For example, the linear coupling coefficients $q_{ps}$ and $q_{sp}$
present nonlinearities which are not directly related to the Kerr
coupling but to the overlapping of the localized nonlinear plasmon
and solution functions $f_{\mathrm{np}}$ and $f_{s}$. 

In order to keep our discussion in the most general form, we have
retained terms that do not appear in the weak coupling approximation,
at least to leading order. If we keep just leading terms in this approximation,
the previous coefficients considerably simplify. Let us briefly recall
the vanishing terms in this approximation:
\begin{itemize}
\item $\delta_{ps}\delta_{sp},\,\,\delta_{ps}\Delta_{sp},\,\,\delta_{sp}\Delta_{ps},\,\,\delta_{ps}\Delta_{K},\,\,\delta_{sp}\Delta_{K}\sim O(e^{-2\kappa_{s}a})\rightarrow0.$
This is so because all coefficients involved in these quadratic products
correspond to overlapping functions of $f_{s}$, and, therefore are
$O(e^{-\kappa_{s}a})$.
\item Besides, as proven before, $\triangle_{ss}\sim O(e^{-2\kappa_{s}a})\rightarrow0$
and $\triangle_{pp}\sim O(e^{-2\kappa_{p}d})\rightarrow0$.
\item And, finally, also proven before, $\triangle'_{\mathrm{K}}\sim O(e^{-2\kappa_{s}a})\rightarrow0.$
\end{itemize}
With all these approximations in mind we find that:
\begin{eqnarray*}
B_{\mathrm{np}}^{2} & \approx & \beta_{\mathrm{np}}^{2}\\
B_{s}^{2} & \approx & \beta_{s}^{2}
\end{eqnarray*}
and
\begin{eqnarray*}
q_{ps} & \approx & \frac{\ensuremath{\Delta}_{ps}}{N_{\mathrm{np}}}\\
q_{sp} & \approx & \frac{\ensuremath{\Delta}_{sp}}{N_{s}},
\end{eqnarray*}
as well as
\begin{eqnarray*}
q_{\mathrm{K}} & \approx & \frac{\Delta_{\mathrm{K}}}{N_{\mathrm{np}}}\\
q'_{\mathrm{K}} & \approx & 0.
\end{eqnarray*}
Summarizing, the general form of the dynamic equations for the soliplasmon
variational parameters in the leading order of the weak coupling approximation
is given by
\begin{eqnarray}
\left(\frac{d^{2}}{dz^{2}}+\beta_{\mathrm{np}}^{2}\right)A & =- & \frac{\ensuremath{\Delta}_{ps}}{N_{\mathrm{np}}}C-\frac{\Delta_{\mathrm{K}}}{N_{\mathrm{np}}}\left(2|A|^{2}C+A^{2}C^{*}\right)\nonumber \\
\left(\frac{d^{2}}{dz^{2}}+\beta_{s}^{2}\right)C & =- & \frac{\ensuremath{\Delta}_{sp}}{N_{s}}A.\label{eq:variational_eq_soliplasmon_parameters_weak_coupling}
\end{eqnarray}

\section{Cases of interest}

The equations found for the evolution of the variational parameters
$A(z)$ and $C(z)$ are, in principle, not restricted to the specific
linear profiles of a MDK structure. Despite we have used the specific
form of $\varepsilon_{L}$ for a MDK structure in some previous steps
to justify some approximations, this procedure has been adopted more
for clarifying purposes than for necessity. In fact, it is not difficult
to realize that the form of the equations and coefficients is preserved
if we assume two general, localized, \emph{linear} dielectric function
profiles for the plasmon ---$\varepsilon_{p}(x)$--- and soliton ---$\varepsilon_{s}(x)$---
fields.

However, since MDK and MK structures are the main physical motivation
of the current study, in this section, we will particularize the general
approach to these two cases of interest. We will work, in principle,
at leading order of the weak coupling approximation, so that Eq.(\ref{eq:variational_eq_soliplasmon_parameters_weak_coupling})
will be our starting point in this section.

\subsection{Coupling of a linear plasmon and a soliton in a MDK structure}

We will assume here that we work with a MDK structure, as presented
in Fig.\ref{fig:MDK}, characterized mathematically by the function
$\varepsilon_{L}(x)$ as presented in Section \ref{sec:variational_equation_for_NL_plasmon}
Eq.(\ref{eq:dielectric_function_MDK}). The plasmon dielectric function
$\varepsilon_{p}(x)$ will be that of a MD structure, as given by
Eq.(\ref{eq:dielectric_function_MD_interface}). On the other hand,
we will assume that Kerr nonlinearities do not affect the plasmon
component, something that can be realistically realized by suitable
playing with the width of the dielectric layer $d$ together with
the amplitude of the SPP field. Therefore, considering the plasmon
component to be linear implies that (since $|A|\ll1$)
\[
\beta_{\mathrm{np}}^{2}\approx\beta_{p}^{2},
\]
as well as suppressing the nonlinear coupling term $O(A^{2})$ in
Eq.(\ref{eq:variational_eq_soliplasmon_parameters_weak_coupling})
because 
\[
\triangle_{\mathrm{K}}\sim\gamma\int_{d}^{\infty}f_{p}(0)f_{\mathrm{p}}^{2}(z)f_{s}(z)\sim\gamma\int_{d}^{\infty}e^{-3\kappa_{p}x}f_{s}\overset{a\gg d}{\sim}O(e^{-3\kappa_{p}d})\rightarrow0.
\]
Hence, we get
\begin{eqnarray}
\left(\frac{d^{2}}{dz^{2}}+\beta_{p}^{2}\right)A & =- & \frac{\ensuremath{\Delta}_{ps}}{N_{p}}C\nonumber \\
\left(\frac{d^{2}}{dz^{2}}+\beta_{s}^{2}\right)C & =- & \frac{\ensuremath{\Delta}_{sp}}{N_{s}}A.\label{eq:variational_eq_soliplasmon_parameters_weak_coupling-1}
\end{eqnarray}
It is possible to obtain explicit expressions for $\Delta_{ps}$ and
$\Delta_{sp}$ in the MDK case. This is possible because the functions
$f_{p}$ and $f_{s}$ are known explicitly. According to the definition
given in Section \ref{sec:variational_equation_for_NL_plasmon}, the
plasmonic function $f_{p}$ is normalized by the peak value of the
$x$ component of the \emph{linear} SPP electric field $\bar{E}_{px}$.
Since we are dealing with the linear solution, we have an analytical
expression for it \cite{Maier_book_2007,Pitarke_RPP_2007}:
\[
\bar{E}_{px}(x)=\begin{cases}
\frac{\beta_{p}E_{0}}{k_{0}\varepsilon_{m}}e^{\kappa_{m}x} & x\le0\\
\frac{\beta_{p}E_{0}}{k_{0}\varepsilon_{d}}e^{-\kappa_{d}x} & x>0
\end{cases}
\]
 The peak value of the SPP field is achieved at $x=0$, so that
\[
f_{p}(x)\equiv\frac{\bar{E}_{px}(x)}{\bar{E}_{px}(0)}=\begin{cases}
e^{\kappa_{m}x} & x\le0\\
e^{-\kappa_{d}x} & x>0.
\end{cases}
\]
Analogously, the $f_{s}$ function is given by $f_{s}(x)=\mathrm{sech}\left[\kappa_{s}\left(x-a\right)\right]$,
where $\kappa_{s}=\left(\gamma/2\right)^{1/2}k_{0}|C|$. We can approximate
the \emph{sech} function by its exponential tail when the overlapping
with the $f_{p}$ function is small, hypothesis which is justified
in the weak coupling approximation. So that, in the overlapping region
\[
\, f_{s}(x)\approx2e^{-\kappa_{s}(a-x)},\,\,\, x<d,\,\,\, a>d>0.
\]
Now, according to their definitions in Eqs(\ref{eq:soliton_projection_integral})
and (\ref{eq:plasmon_projection_integrals}) and to the form of the
local variations of the linear dielectric functions $\triangle\varepsilon_{p}$
and $\triangle\varepsilon_{s}$ in Eqs.(\ref{eq:local_variation_dielectric_function_plasmon})
and (\ref{eq:local_variation_dielectric_function_soliton}), we obtain
for the coupling coefficient
\begin{eqnarray}
\Delta_{ps} & = & k_{0}^{2}\left[\left(\varepsilon_{m}-\varepsilon_{\mathrm{K}}\right)\int_{-\infty}^{0}f_{p}f_{s}+\left(\varepsilon_{d}-\varepsilon_{\mathrm{K}}\right)\int_{0}^{d}f_{p}f_{s}\right]\nonumber \\
 & \overset{d\ll\left|\kappa_{d}-\kappa_{s}\right|^{-1}}{\approx} & 2e^{-\kappa_{s}a}k_{0}^{2}\left[\frac{\varepsilon_{m}-\varepsilon_{\mathrm{K}}}{\kappa_{s}+\kappa_{m}}+\left(\varepsilon_{d}-\varepsilon_{\mathrm{K}}\right)d\right],\label{eq:explicit_eq_Delta_ps}
\end{eqnarray}
where in the last step we show an expression valid for sufficiently
thin dielectric slabs. 

For $\Delta_{sp}$ we have
\begin{eqnarray}
\Delta_{sp} & = & k_{0}^{2}\left(\varepsilon_{\mathrm{K}}-\varepsilon_{d}\right)\int_{d}^{\infty}f_{s}f_{p}\nonumber \\
 & \approx & 2k_{0}^{2}\left(\frac{\varepsilon_{\mathrm{K}}-\varepsilon_{d}}{\kappa_{d}-\kappa_{s}}\right)e^{-d\kappa_{d}+(d-a)\kappa_{s}},\label{eq:explicit_eq_Delta_sp}
\end{eqnarray}
where in order to give the approximated expression above we had to
assume that $\kappa_{d}>\kappa_{s}$ and $a\gg d$, so we could approximate
$f_{s}$ by its right-hand-side exponential tail.

In the same way, the plasmonic linear ``norm'' $N_{p}=\int_{\mathbb{R}}f_{p}^{2}$
can be evaluated to give
\begin{equation}
N_{p}=\frac{1}{2}\left(\frac{1}{\kappa_{d}}+\frac{1}{\kappa_{m}}\right),\label{eq:Np}
\end{equation}
whereas the solitonic one $N_{s}=\int_{\mathbb{R}}f_{s}^{2}$ takes
the simple expression
\begin{equation}
N_{s}=\frac{2}{\kappa_{s}}.\label{eq:Ns}
\end{equation}
All the evolution equations we have obtained up to now are second
order in the derivative with respect to the propagation variable $z$.
Physically speaking, they are non-paraxial. However, the physical
configuration under consideration, in which the soliton propagates
in parallel to the metal/dielectric interface (see Fig.\ref{fig:MDK}
) following the $z$ axis, exhibits a clear paraxial character. This
fact indicates that a slowly varying approximation for the plasmon
and soliton components with respect the propagation parameter $z$
is expected to be adequate for the analysis of this case as well.
Certainly, it will properly describe propagation in the regions where
most of the energy is localized, namely, those where the plasmon and
soliton components evolve. Since we are working in the weak coupling
approximation the plasmon and soliton regions are, by construction,
clearly distinguishable in the form of weakly coupled plasmon and
soliton modes propagating along the $z$ axis. However, despite these
components are essentially paraxial, the energy exchange between them
induced by the coupling is not necessarily so. It can include, in
principle, non-paraxial components associated to a very fast, in the
sense of rapidly varying in $z$, energy exchange %
\footnote{This exchange of energy is visible in the figures presented in Ref.\cite{Milian_OL_2012}
showing the propagation of a perturbed soliplasmon field along the
surface, in which the flux of the Poynting vector is represented.
One should remark at this point that these simulations are the result
of solving the \emph{full} nonlinear vector Maxwell's equations (\ref{eq:NL_vector_wave_eq})
numerically.%
}. Thus, we expect that, with the exception of this type of rapid oscillations,
the slowly varying approximation will provide also a good approximation
to the solution. Thus, we introduce the slowly varying plasmonic $\widetilde{A}(z)$
and soliton $\widetilde{C}(z)$ envelops in the usual way, taking
$n_{\mathrm{K}}=\varepsilon_{\mathrm{K}}^{1/2}$ as the reference
index,
\begin{eqnarray*}
A(z) & = & \widetilde{A}(z)e^{ik_{0}n_{\mathrm{K}}z}\\
C(z) & = & \widetilde{C}(z)e^{ik_{0}n_{\mathrm{K}}z},
\end{eqnarray*}
so that, we can write (after neglecting $\left|d^{2}\widetilde{A}/dz^{2}\right|\ll ik_{0}n_{k}\left|d\widetilde{A}/dz\right|$,
idem for $\widetilde{C}$),
%
%
%
\begin{eqnarray*}
-i\frac{d\tilde{A}}{dz} & = & \mu_{p}\tilde{A}+q\tilde{C}\\
-i\frac{d\tilde{C}}{dz} & = & \mu_{s}\tilde{C}+\bar{q}\tilde{A},
\end{eqnarray*}
in which we have defined the paraxial propagation constants $\mu_{p}$
and $\mu_{s}$ as
\begin{eqnarray*}
\mu_{p} & \equiv & \frac{\left(\beta_{p}^{2}-k_{0}^{2}\varepsilon_{\mathrm{K}}\right)}{2k_{0}n_{\mathrm{K}}}\\
\mu_{s} & \equiv & \frac{\left(\beta_{s}^{2}-k_{0}^{2}\varepsilon_{\mathrm{K}}\right)}{2k_{0}n_{\mathrm{K}}}=\frac{k_0 \gamma}{4n_{\mathrm{K}}}|\tilde{C}|^{2}.
\end{eqnarray*}
In the last equation we have used the explicit expression for the
Helmholtz soliton propagation constant $\beta_{s}=k_{0}\left(\varepsilon_{\mathrm{K}}+\frac{\gamma}{2}|C|^{2}\right)^{1/2}$,
as introduced in Section \ref{sec:variational_eq_soliton}. 
%
%

Analogously, we have defined the paraxial coupling coefficients $q$
and $\bar{q}$ as
\begin{eqnarray*}
q & \equiv & \frac{\Delta_{ps}}{2k_{0}n_{\mathrm{K}}N_{p}}=\frac{k_{0}}{2n_{\mathrm{K}}N_{p}}\int_{\mathbb{R}}f_{p}\triangle\varepsilon_{s}f_{s}\\
\bar{q} & \equiv & \frac{\Delta_{sp}}{2k_{0}n_{\mathrm{K}}N_{s}}=\frac{k_{0}}{2n_{\mathrm{K}}N_{s}}\int_{\mathbb{R}}f_{s}\triangle\varepsilon_{p}f_{p}.
\end{eqnarray*}
Explicit expressions for the paraxial couplings can be given in the
same way as for their non-paraxial counterparts in the case of a MDK
structure. From expressions (\ref{eq:explicit_eq_Delta_ps}) and (\ref{eq:explicit_eq_Delta_sp})
for $\Delta_{ps}$ and $\Delta_{sp}$, we have
\begin{eqnarray*}
q & = & \frac{k_{0}}{n_{\mathrm{K}}N_{p}}e^{-\kappa_{s}a}\left(\varepsilon_{m}-\varepsilon_{\mathrm{K}}\right)\left[\frac{1}{\kappa_{m}+\kappa_{s}}+\left(\frac{\varepsilon_{d}-\varepsilon_{\mathrm{K}}}{\varepsilon_{m}-\varepsilon_{\mathrm{K}}}\right)d\right]\\
\bar{q} & = & \frac{k_{0}}{n_{\mathrm{K}}N_{s}}e^{-\kappa_{s}a}\left(\varepsilon_{\mathrm{K}}-\varepsilon_{d}\right)\left[\frac{1}{\kappa_{d}-\kappa_{s}}-d\right]
\label{q_paraxial}
\end{eqnarray*}
An important consequence of the previous analysis is the fact that
the plasmon-soliton coupling is asymmetric since, in general, $q\neq\bar{q}$.
The previous expressions allow us to obtain relevant information about
the characteristics of the coupling ratio $\bar{q}/q$. At this point,
we are only interested in estimating the order of magnitude of this
ratio, so that if we consider only the leading terms in $d$, the explicit expressions for $N_p$ and $N_s$, and the relation between
the inverse penetration lengths in the metal and dielectric,
$\kappa_{m}$ and $\kappa_{d}$, and the MD dielectric constants \cite{Pitarke_RPP_2007},
\begin{eqnarray*}
\kappa_{m} & = & -k_{0}\varepsilon_{m}\sqrt{\frac{-1}{\varepsilon_{m}+\varepsilon_{d}}}\approx k_{0}\left(-\varepsilon_{m}\right)^{1/2}\\
\kappa_{d} & = & k_{0}\varepsilon_{d}\sqrt{\frac{-1}{\varepsilon_{m}+\varepsilon_{d}}}\approx k_{0}\varepsilon_{d}\left(-\varepsilon_{m}\right)^{-1/2},
\end{eqnarray*}
we can simplify the ratio into 
\begin{equation}
\frac{\bar{q}}{q}\approx\frac{\kappa_{s}}{4\kappa_{d}}\frac{\kappa_{m}\left(\varepsilon_{d}-\varepsilon_{\mathrm{K}}\right)}{\kappa_{d}\left(-\varepsilon_{m}\right)}=\frac{\kappa_{s}}{\kappa_{d}}\left(\frac{\varepsilon_{d}-\varepsilon_{\mathrm{K}}}{4\varepsilon_{d}}\right),\label{eq:ratio_qbar_q}
\end{equation}
where we have simultaneously assumed that $\kappa_{m}\gg\kappa_{d}\gg\kappa_{s}$
and $\left|\varepsilon_{m}\right|\gg\varepsilon_{d},\varepsilon_{\mathrm{K}}$. The latter approximation is pretty realistic for common MD interfaces. The former one simply indicates that the plasmon penetration length in the metal is significantly smaller than in the dielectric (which is consistent with the previous approximation since $\kappa_m/\kappa_d = \varepsilon_m/\varepsilon_d \gg 1$) whereas the plasmon penetration length in the dielectric is, in turn, smaller than the typical spatial soliton width (which is also a reasonable assumption for paraxial solitons).

The previous result shows that this ratio is generally small by two
reasons: first, and most important, in this regime $\kappa_{s}/\kappa_{d}\ll1$,
and, second, for standard dielectric and Kerr materials the dielectric
constants ratio (term in parentheses in Eq.(\ref{eq:ratio_qbar_q}))
can be also small. Let us be more precise and introduce explicit expressions for $\kappa_{d}$
and $\kappa_{s}$. For the soliton inverse penetration length $\kappa_{s}=\left(\gamma/2\right)^{1/2}k_{0}|C_{0}|$
we choose the peak amplitude to be the initial one since, in order
to preserve the weak coupling approximation, it must be true that
$\kappa_{s}(z)\sim\kappa_{s}(0)$ for all values of $z$. It is useful
to introduce the dimensionless nonlinear coefficient $ $ $\bar{\gamma}=\gamma|C_{0}|^{2}$,
normalized to the initial soliton peak amplitude. In this way, we
get
\[
\frac{\bar{q}}{q}\approx\left(\frac{\varepsilon_{d}-\varepsilon_{\mathrm{K}}}{4\varepsilon_{d}^{2}}\right)\left(\frac{-\varepsilon_{m}\bar{\gamma}}{2}\right)^{1/2}.
\]
If we introduce some typical numbers for the dielectric constants,
say $\varepsilon_{d}=1.5^{2}$ , $\varepsilon_{\mathrm{K}}=2^{2}$
and $\varepsilon_{m}=-80$, we can make the following estimation
\[
\frac{\bar{q}}{q}\sim10^{-1}\left(40\bar{\gamma}\right)^{1/2}\sim0.5\bar{\gamma}^{1/2}.
\]
This ratio is small because in ordinary cases $\bar{\gamma}\ll1$.
Now, for a standard nonlinear Kerr medium with a nonlinear index $n_{2}\sim10^{-19}\mathrm{(m^{2}/W)}$
(one order of magnitude higher than that of silica), corresponding
to a value of $\gamma=c\epsilon_{0}\varepsilon_{\mathrm{K}}n_{2}\text{\ensuremath{\sim}}10^{-22}\mathrm{(m^{2}/V^{2})}$,
and a peak electric field $C_{0}=E_{0}\sim10^{7}\left(\mathrm{V/m}\right)$,
which would be a typical value for a $10\,\mathrm{\mu m}$ wide (along $x$ direction), $100\,\mu\mathrm{m}$
(or larger) high beam (along $y$ direction) ---as an approximation to a 1D soliton---, with peak
power of $P_{0}\gtrsim10\,\mathrm{kW}$, we would get $\bar{\gamma}\sim10^{-5}$.
The order of magnitude of the coupling ratio would be then 
 $\bar{q}/q\sim10^{-3}$ which shows how small this ratio can be
in typical situations. 

In summary, the paraxial equations for a soliplasmon state in a MDK
system in which the plasmon component behaves linearly are, under
all the approximations carefully explained in the present section
and in matrix form,
\begin{eqnarray*}
-i\frac{d}{dz}\left(\begin{array}{c}
\tilde{A}\\
\tilde{C}
\end{array}\right) & = & \left(\begin{array}{cc}
\mu_{p} & q\\
\bar{q} & \mu_{s}
\end{array}\right)\left(\begin{array}{c}
\tilde{A}\\
\tilde{C}
\end{array}\right),
\end{eqnarray*}
in which, in typical conditions, the soliton-to-plasmon coupling
$q$ is much stronger that the plasmon-to-soliton one $\bar{q}$:
$\bar{q}\ll q$. This equation is the one presented in Ref.\cite{Milian_OL_2012}
in which $c_{p}=\tilde{A}$ and $c_{s}=\tilde{C}$ are the paraxial
plasmon and soliton envelopes, respectively.

\begin{figure}
\hfill{}\includegraphics[scale=0.4]{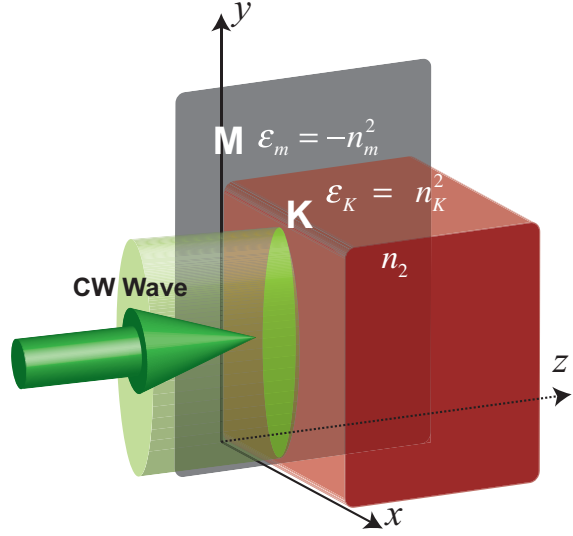}\hfill{}

\caption{Parallel illumination of a metal/Kerr interface from the Kerr medium.\label{fig:MK}}
\end{figure}

\subsection{Coupling of a nonlinear plasmon and a soliton in a MK structure}

We analyze now another interesting case, namely, that of a metal surface
directly attached to the Kerr medium and subject to parallel illumination
by means of a spatial soliton of the Kerr medium (see Fig.\ref{fig:MK}).
In this situation, the dielectric slab of the MDK structure analyzed
in the previous section is no longer present. This has two important
consequences in the mathematical description of the problem: firstly,
the linear index contrast between the dielectric and Kerr medium of
the MDK structure disappears, $\varepsilon_{d}=\varepsilon_{\mathrm{K}}$;
and, secondly, because the Kerr medium is now attached to the metal,
Kerr nonlinearities directly affect the SPP, so that it is required
to consider a nonlinear plasmon instead of its linear counterpart.
Let us see how these two features affect the soliplasmon propagation
equation.

As before, we work at leading order of the weak coupling approximation,
so Eq.(\ref{eq:variational_eq_soliplasmon_parameters_weak_coupling})
is the correct variational propagation equation to use in this case.
The first consequence of working with a MK structure instead of with
a MDK one is that the plasmon-to-soliton coupling completely disappears.
Indeed, according to Eq.(\ref{eq:local_variation_dielectric_function_plasmon})
$\Delta\varepsilon_{p}=0$ and, therefore, $\triangle_{sp}=0$ according
to the definition (\ref{eq:soliton_projection_integral}). On the
other hand, the plasmon propagation constant $\mathrm{\beta{}_{np}}$
is now that of a stationary nonlinear plasmon and it depends on the
SPP amplitude $|A|$. The analytic form of $\beta_{\mathrm{np}}$
on $|A|$ is know only in some cases \cite{Agranovich_JETP_1980}. In the
general case, $\beta_{\mathrm{np}}^{2}$ is nothing but the eigenvalue
of the nonlinear equation for the electric components of the nonlinear
SPP field (\ref{eq:NL_equation_only_transverse}). As explained in
Section \ref{sec:variational_equation_for_NL_plasmon}, the nonlinear
terms in this equation depend nonlinearly on the plasmon amplitude
as $|A|^{2}$ . So that, $\beta_{\mathrm{np}}^{2}=\beta_{\mathrm{np}}^{2}(|A|^{2})$
and we can always perform a Taylor expansion in $|A|^{2}$:
\[
\beta_{\mathrm{np}}^{2}=\beta_{\mathrm{p}}^{2}+k_{0}^{2}\gamma_{p}|A|^{2}+O(|A|^{4}).
\]
We are interested here in the first order nonlinear corrections to
the linear case, so that we keep the first correction only and neglect
$O(|A|^{4})$ terms. Consequently, to leading order in the weak coupling
approximation and to first order in $|A|^{2}$, we have 
\begin{eqnarray}
\left(\frac{d^{2}}{dz^{2}}+\beta_{\mathrm{np}}^{2}\right)A & =- & \frac{\ensuremath{\Delta}_{ps}}{N_{\mathrm{np}}}C-\frac{\Delta_{\mathrm{K}}}{N_{\mathrm{np}}}\left(2|A|^{2}C+A^{2}C^{*}\right)\nonumber \\
\left(\frac{d^{2}}{dz^{2}}+\beta_{s}^{2}\right)C & = & 0.
\label{eq:variational_eq_soliplasmon_MK}
\end{eqnarray}
We can write the previous equation in a form that resemble that of
a linear plasmon in Eq.(\ref{eq:variational_eq_soliplasmon_parameters_weak_coupling-1}).
\begin{eqnarray}
\left[\frac{d^{2}}{dz^{2}}+\bar{\beta}_{np}^{2}\left(A,C\right)\right]A & =- & \frac{\ensuremath{\bar{\Delta}}_{ps}\left(|A|,|C|\right)}{N_{\mathrm{np}}}C\nonumber \\
\left[\frac{d^{2}}{dz^{2}}+\beta_{s}^{2}\left(|C|\right)\right]C & = & 0,\label{eq:variational_eq_soliplasmon_parameters_MK_bis}
\end{eqnarray}
where now
\begin{eqnarray}
\bar{\beta}_{np}^{2}\left(A,C\right) & = & \beta_{\mathrm{p}}^{2}+k_{0}^{2}\gamma_{p}|A|^{2}+\frac{\Delta_{\mathrm{K}}\left(|C|\right)}{N_{\mathrm{np}}}AC^{*}\nonumber \\
\ensuremath{\bar{\Delta}}_{ps}\left(|A|,|C|\right) & = & \ensuremath{\Delta}_{ps}(|C|)+2\Delta_{\mathrm{K}}(|C|)|A|^{2}.\label{eq:renormalized_beta_delta_NL_plasmon}
\end{eqnarray}
Note that the effective propagation constant is now a complex number
$\bar{\beta}_{\mathrm{np}}\in\mathbb{C}$ because of the nonlinear
$AC^{*}$ coupling. Thus the nonlinear soliton-to-plasmon coupling
simultaneously affects both the phase velocity of the nonlinear plasmon
and the plasmon amplitude, due to the presence of a nonzero\emph{
gain-loss} coefficient proportional to the imaginary part of the effective
propagation constant: $\Im(\bar{\beta}_{\mathrm{np}}^{2})\sim\sin\left(\varphi_{p}-\varphi_{s}\right)\neq0,$
in general. There is also a nonlinear plasmonic modification of the
coupling coefficient $\Delta_{ps}$ in Eq.(\ref{eq:renormalized_beta_delta_NL_plasmon})
that can slightly modify the nature of the coupling with respect the
pure MDK case. These two effects are \emph{additional} to those associated
to the $A$-independent coefficient $ $$\Delta_{ps}(|C|)$ appearing
in the case of the linear-plasmon/soliton coupling in a MDK structure
\cite{Milian_OL_2012}. 

We see that, at this order of the weak coupling approximation, the
soliton equation decouples from the plasmon one. Physically speaking,
the soliton acts as a non-depleting reservoir pumping the non-linear
plasmon without experience any energy exchange with the SPP. Obviously,
this is only true within the order of our approximation because we
have truncated higher-order terms. Indeed, there exists this type
of energy exchange from the plasmon to the soliton but this has to
fulfill two conditions: (i), it has to be very small, i.e., in mathematical
terms is has to be at least $O(e^{-\kappa_{s}a})^{2}$ since this
is the order of terms neglected in our weak coupling approximation;
and (ii), it necessarily has to arise from terms neglected in the
process of deriving the variational equations at leading order, since
the leading order plasmon-to-soliton coupling for a MK structure,
linked to variations in the linear dielectric functions, is strictly
zero inasmuch $\triangle\varepsilon_{p}=0$. So, we would need to
go to next-to-leading order in the weak coupling approximation if
we wanted to account for these effects. 

This generalization to next-to-leading order is certainly more complicated
because we need to include terms that we have neglected in previous
analysis. If we keep all the approximations to the same order, as
before, and we only go to the next order in the weak coupling approximation,
we need to consider the four terms in the Kerr coupling $Q_{\mathrm{K}}$
defined in Eq.(\ref{eq:general_Kerr_coupling}) . We have to retained
now the first two terms. We neglected them before because they provided
terms $O(e^{-\kappa_{s}a})^{2}$ that we want to keep now. The plasmon
and soliton projections provide two more terms linked to them. The
plasmonic projection (\ref{eq:plasmonic_projection}) is now:
\begin{eqnarray*}
N_{\mathrm{np}}\left[\frac{d^{2}}{dz^{2}}A+\beta_{\mathrm{np}}^{2}A\right]+\delta_{ps}\left[\frac{d^{2}}{dz^{2}}C+\beta_{\mathrm{s}}^{2}C\right] & = & -\triangle_{pp}A-\triangle_{ps}C\\
 &  & \hspace{-3cm}-\triangle_{\mathrm{K}}\left[2|A|^{2}C+A^{2}C^{*}\right]-\Gamma_{\mathrm{K}}\left[2|C|^{2}A+C^{2}A^{*}\right],
\end{eqnarray*}
whereas the soliton projection reads (\ref{eq:soliton_projection})
\begin{eqnarray*}
\delta_{sp}\left[\frac{d^{2}}{dz^{2}}A+\beta_{\mathrm{np}}^{2}A\right]+N_{s}\left[\frac{d^{2}}{dz^{2}}C+\beta_{\mathrm{s}}^{2}C\right] & = & -\triangle_{ss}C-\triangle_{sp}A\\
 &  & \hspace{-3cm}-\triangle'_{\mathrm{K}}\left[2|A|^{2}C+A^{2}C^{*}\right]-\Gamma'_{\mathrm{K}}\left[2|C|^{2}A+C^{2}A^{*}\right],
\end{eqnarray*}
where we have introduced two new Kerr coupling coefficients:
\begin{eqnarray*}
\Gamma_{\mathrm{K}} & \equiv & k_{0}^{2}\int_{\mathbb{R}}\gamma f_{\mathrm{np}}(0)f_{\mathrm{np}}(z)f_{s}^{2}(z)\\
\Gamma'_{\mathrm{K}} & \equiv & k_{0}^{2}\int_{\mathbb{R}}\gamma f_{s}(0)f_{\mathrm{np}}(z)f_{s}^{2}(z).
\end{eqnarray*}
However, since $\triangle\varepsilon_{p}=0$, we know that their two
projections vanish, so that, $\Delta_{sp}=0$ and $\Delta_{pp}=0$.
If we follow now the demonstration given in Section \ref{sec:variational_eq_for_soliplasmon},
we can conclude that the form of the generalized equation (\ref{eq:variational_eq_soliplasmon_parameters})
has to be substituted by a new one including the new Kerr-coupling
terms and in which $q_{sp}=0$ (from its definition Eq.(\ref{eq:q_linear_coefficients})
along with $\Delta_{sp}=0$ and $\Delta_{pp}=0$):
\begin{eqnarray}
\left(\frac{d^{2}}{dz^{2}}+B_{\mathrm{np}}^{2}\right)A & = & -q_{ps}C-q_{\mathrm{K}}\left(2|A|^{2}C+A^{2}C^{*}\right)-p_{\mathrm{K}}\left(2|C|^{2}A+C^{2}A^{*}\right)\nonumber \\
\left(\frac{d^{2}}{dz^{2}}+B_{s}^{2}\right)C & = & -q'_{\mathrm{K}}\left(2|A|^{2}C+A^{2}C^{*}\right)-p'_{\mathrm{K}}\left(2|C|^{2}A+C^{2}A^{*}\right),\label{eq:variational_eq_soliplasmon_parameters_general}
\end{eqnarray}
where we have introduced the new nonlinear coefficients
\begin{eqnarray}
p{}_{K} & = & \frac{N_{s}\Gamma{}_{\mathrm{K}}-\delta_{ps}\Gamma'_{\mathrm{K}}}{N_{\mathrm{np}}N_{s}-\text{\ensuremath{\delta}}_{ps}\ensuremath{\delta_{sp}}}\nonumber \\
p'_{K} & = & \frac{N_{\mathrm{np}}\Gamma'_{\mathrm{K}}-\delta_{sp}\Gamma_{\mathrm{K}}}{N_{\mathrm{np}}N_{s}-\text{\ensuremath{\delta}}_{ps}\ensuremath{\delta_{sp}}}.\label{eq:p_K_coefficients}
\end{eqnarray}
We see that, as expected, despite there is no trace of \emph{linear}
plasmon-to-soliton coupling, nonlinear coupling occurs in the next-to-leading
order weak coupling approximation. It is revealing to consider the
case of a very weak plasmon field, in which we are approaching the
linear plasmon limit and thus terms $ $$O(A^{2})$ can be neglected
in Eq.(\ref{eq:variational_eq_soliplasmon_parameters_general}):
\begin{eqnarray}
\left(\frac{d^{2}}{dz^{2}}+B_{\mathrm{np}}^{2}\right)A & = & -q_{ps}C-p_{\mathrm{K}}\left(2|C|^{2}A+C^{2}A^{*}\right)\nonumber \\
\left(\frac{d^{2}}{dz^{2}}+B_{s}^{2}\right)C & = & -p'_{\mathrm{K}}\left(2|C|^{2}A+C^{2}A^{*}\right).\label{eq:variational_eq_soliplasmon_parameters_small_A}
\end{eqnarray}
The previous equation makes clear that there is a second mechanism
for plasmon-to-soliton coupling absent at leading order. Soliton can
be also pumped or depleted by the plasmon field through a mechanism
based on the Kerr coupling terms in Eq.(\ref{eq:variational_eq_soliplasmon_parameters_small_A}).
These terms can be understood as nonlinear sources generating gain
or loss on the soliton parameter $C$ and driven by the plasmon field.
As before, this mechanism can be visualized by renormalizing the soliton
propagation constant in the following way:

\[
\bar{B}_{s}^{2}\equiv B_{s}^{2}+p'_{\mathrm{K}}\left(2C^{*}A+CA^{*}\right),
\]
which permits to write the soliton equation as:
\[
\left(\frac{d^{2}}{dz^{2}}+\bar{B}_{s}^{2}\right)C=0.
\]
Analogously as in a previous analysis, the imaginary part of the renormalized
propagation constant give us essential information about this nonlinear
mechanism 
\[
\Im(\bar{B}_{s}^{2})=p'_{\mathrm{K}}|C||A|\sin\left(\varphi_{p}-\varphi_{s}\right),
\]
which indicates that this type of term generates nonlinear gain or
loss depending on the value of the soliplasmon relative phase.

Another way of understanding the physical nature of the plasmon-to-soliton
nonlinear term (the one proportional to $p'_{\mathrm{K}}$) in Eq.(\ref{eq:variational_eq_soliplasmon_parameters_small_A}),
and more specifically the one depending on $|C|^{2}$, is by considering
that its origin is the \emph{nonlinear} modulation of the dielectric
function in the Kerr medium. We know that the absence of \emph{linear}
modulation ($\triangle\varepsilon_{p}=0$) makes the plasmon-to-soliton
linear coupling to vanish. However, even if there is no linear modulation
in the region where the soliton is localized (we assume an homogeneous
nonlinear dielectric medium), the Kerr effect does induce a nonlinear
modulation of the refractive index or, equivalently, of the dielectric
function $\varepsilon(x)=\varepsilon_{\mathrm{K}}+\gamma|E_{s}(x)|^{2}$.
Therefore, there exists, in fact, an effective modulation of $\varepsilon(x)$
induced by the Kerr nonlinearity that, in turn, originates a local
variation of the dielectric function for the plasmon field, as defined
in Eq.(\ref{eq:local_variation_dielectric_function_plasmon}), given
by 
\[
\triangle\varepsilon_{p}^{\mathrm{NL}}(x)=\varepsilon(x)-\varepsilon_{p}(x)=\varepsilon(x)-\varepsilon_{\mathrm{K}}=\gamma|E_{s}|^{2}\,\,\,\mathrm{if}\,\,\, x>0.
\]
Since this term is nonzero, an effective plasmon-to-soliton variational
coupling, analogous to the linear one $\Delta_{sp}$ (Eq.(\ref{eq:soliton_projection_integral})),
is expected (recall that the linear coefficient vanish, $\Delta_{sp}=0$)
\[
\triangle_{sp}^{\mathrm{NL}}\equiv k_{0}^{2}\int_{\mathbb{R}}f_{s}(0)\triangle\varepsilon_{p}^{\mathrm{NL}}f_{\mathrm{np}}(z)=k_{0}^{2}|C|^{2}\int_{\mathbb{R}}\gamma f_{s}(0)f_{s}^{2}(z)f_{\mathrm{np}}(z)=\Gamma'_{\mathrm{K}}|C|^{2}.
\]
The previous variational term corresponds to the soliton projection
of $\triangle\varepsilon_{p}^{\mathrm{NL}}$. One recognizes that
this term is responsible of the first term of the $p'_{\mathrm{K}}$
coefficient in Eq.(\ref{eq:p_K_coefficients}). This term can be interpreted
as the leading order effect of soliton-soliton interaction in the
weak coupling regime. In this case, the nonlinear plasmon, acting
as a soliton, couples to the soliton tail though its own asymptotic
exponential tail by means of the Kerr term. The soliton-soliton interaction
exists independently of the existence of a linear modulation and,
therefore, occurs even in a completely homogeneous medium. For this
reason, it is expected to exist even if the soliton moves in a completely
homogeneous medium, as in the present case in which we deal with a
MK structure. Analogously, the nonlinear self-modulated local variation
of the dielectric function $\triangle\varepsilon_{p}^{\mathrm{NL}}$
generates also a next-to-leading order coupling through its plasmon
projection, which physically can be interpreted as generated by a
nonlinearly induced term $\Delta_{pp}^{\mathrm{NL}}$ of the type
given in Eq.(\ref{eq:Delta_pp_coefficient}) (recall that there is
no linear counterpart of this term since $\Delta_{pp}=0$):
\[
\triangle_{pp}^{\mathrm{NL}}\equiv k_{0}^{2}\int_{\mathbb{R}}f_{\mathrm{np}}(0)\triangle\varepsilon_{p}^{\mathrm{NL}}f_{\mathrm{np}}(z)=k_{0}^{2}|C|^{2}\int_{\mathbb{R}}\gamma f_{\mathrm{np}}(0)f_{s}^{2}(z)f_{\mathrm{np}}(z)=\Gamma{}_{\mathrm{K}}|C|^{2}.
\]
This coupling coefficient generates the second term in the expression
for $p'_{\mathrm{K}}$ (\ref{eq:Delta_pp_coefficient}). 

\section{Conclusions}

We have introduced a variational approach to properly understand the physics behind the problem of the nonlinear excitation of a SPP by a spatial soliton. Unlike in the original proposal in Ref.\cite{Bliokh_PRA_2009}, the resulting nonlinear resonator model is obtained from first principles, which permits, in this way, to provide an approximate  solution of the full vector Maxwell's wave equation (\ref{eq:NL_vector_wave_eq}) for configurations close to our soliplasmon {\em ansatz} (\ref{eq:soliplasmon_ansatz}). In physical terms, variational equations extract the most relevant information of soliplasmon resonances as bound states of a soliton and a linear or nonlinear SPP.  They provide the dynamics of a SPP and a soliton propagating along a metal/dielectric interface in the presence of a continuous exchange of electromagnetic energy between them in a process controlled by a nonlinear coupling constant, which is proportional to the soliton field at the interface $e^{-\kappa_{s}a}$ (see Eq.(\ref{q_paraxial})). Since the inverse soliton penetration length $\kappa_{s}=\left(\gamma/2\right)^{1/2}k_{0}|C(z)|$ depends nonlinearly  on the soliton amplitude, even the simpler case of a soliton bounded to a linear SPP presents special features with respect other well-know nonlinear resonator models \cite{Eks_PRA_2011}. 

On the other hand, the fact that soliplasmon resonances exists even in the presence of a low power SPP component, i.e, a linear SPP, demonstrates that nonlinearities in a soliplasmon play also a supplementary role than the one that is commonly attributed to them in nonlinear plasmonics.  The fact that SPP's can support very high intensities very close to the metal/nonlinear-dielectric interface is the origin of most of the nonlinear effects reported in the literature. They are responsible of generation of second harmonic (see, e.g., \cite{Zayats_PR_2005} and references therein) but also of nonlinearities of the Kerr type. They are known as early as in the 80's \cite{Agranovich_JETP_1980,Tomlinson_OL_1980,Akhmediev_JETP_1982,Yu1983a,Leung_PRB_1985,Mihalache_PS_1984,Stegeman1985a,Ariyasu_JAP_1985}. More recently, substantial advances in the field of plasmonics have refreshed the possibility of exploiting these effects for nanophotonics applications using available technology, so a renovated interest in nonlinear effects induced by Kerr materials interacting with metals have been reflected in the literature (see, for example, \cite{Feigenbaum_OL_2007,Davoyan_OE_2009,Ye_PRL_2010,Salgueiro_APL_2010,Skryabin_JOSAB_2011,Marini_OE_2011,Marini_PRA_2011,Milian_APL_2011,Noskov_PRL_2012}).
The high intensities reached in the dielectric in the vicinity of the metal interface associated to a high-power SPP mode are able to enhance the nonlinear response of a Kerr medium directly attached to the metal. This response can, in return, modify the propagation properties of the SPP. As pointed out as early as in Ref.\cite{Stegeman1985a}, "when nonlinear dielectric media are in contact with a metal surface, the surface plasmon polaritons guided
by that interface become power-dependent." This can be considered a standard definition of a nonlinear plasmon. The fact that the propagation constant of the SPP becomes power-dependent explicitly shows that, according to this definition of a nonlinear plasmon, the latter is continuously connected to the linear SPP in a standard $P$ {\em vs} $\mu$ representation. From this perspective, soliplasmons are not nonlinear plasmons since their resonant nature forces their two branches ---corresponding to 0- and $\pi$- soliplasmon solutions--- to be detached from that of a linear SPP in a  $P$ {\em vs} $\mu$ representation (see its resonant behavior in Ref.\cite{Milian_OL_2012}). Physically, in a soliplasmon, plasmon and soliton preserve their identity as localized solutions even though they are interacting. In a nonlinear plasmon, the soliton cannot be resolved as a second spatially localized component. Our variational approach for a soliplasmon formalizes this features explicitly by assuming the existence of two separately localized plasmon and soliton components in our {\em ansatz} (\ref{eq:general_ansatz}). In this way, our variational equations distinguish between two different types of nonlinear effects: (i) those affecting the {\em uncoupled} propagation of the plasmon and soliton components, i.e, those which appear even when they do not interact; and (ii) those related to the interaction. In the case of the SPP, the former are taken into account in our variational formalism through the dependence of the plasmon propagation constant  $\beta_{\mathrm{np}}^{2}=\beta_{\mathrm{np}}^{2}(|A|^{2})$ on the SPP amplitude. This is analogous to the well-known behavior of the soliton propagation constant with respect to its amplitude: $\beta_{s}^{2}=\beta_{s}^{2}(|C|^{2})$. The second type of nonlinear effects are related to coupling. Here we can, in turn, distinguish two different categories. The first one is related to typical soliton-to-soliton coupling, as those appearing in Eqs.(\ref{eq:variational_eq_soliplasmon_MK}) or (\ref{eq:variational_eq_soliplasmon_parameters_small_A}), which show a form analogous to cross-phase modulation \cite{Kivshar_book_2006}. The second one is related to the modulation of the linear dielectric function and it is the analogous to the coupling between neighboring atoms in a crystal in the so-called {\em tight binding approximation} \cite{Ashcroft_book_1976}. The coupling here is giving by the overlapping of the wave functions of two solutions of individual atomic sites detached from one another weighted by the difference of the total periodic potential with respect the atomic one. This is the nature of the coupling terms $\Delta_{sp}$ and $\Delta_{ps}$  involving the local variations of the linear dielectric function  $\triangle\varepsilon_{p}$ and $\triangle\varepsilon_{s}$ in Eqs.(\ref{eq:plasmon_projection_integrals}) and (\ref{eq:soliton_projection_integral}). This is in fact the dominant coupling term in the weak coupling approximation, as demonstrated by our variational equations for the interaction of a linear plasmon and a soliton (\ref{eq:variational_eq_soliplasmon_parameters_weak_coupling-1}). However, since the overlapping solutions are in this case nonlinear (in this case, that of the the soliton), the coupling becomes itself nonlinear. It is precisely this second type of nonlinearity in the coupling what causes  the soliplasmon variational model to be qualitatively different from other coupled nonlinear systems \cite{Eks_PRA_2011}.





The work of A. F. was partially supported by the MINECO under the TEC2010-15327 grant.

\end{document}